\documentstyle[prd,aps,floats,preprint,epsf]{revtex}

\tightenlines

\newcommand{\be}{\begin{equation}}
\newcommand{\ee}{\end{equation}}
\begin{document}

\preprint{BROWN-HET-1172}

\title{On the Interaction of Monopoles and Domain Walls} 
\author{Stephon Alexander$^1$, Robert Brandenberger$^1$, 
Richard Easther$^1$ and Andrew Sornborger$^2$}
\address{~\\$^1$ Physics Department, Brown University, Providence, RI
02912,  USA; 
~\\$^2$ NASA/Fermilab Astrophysics Group, Fermi National Accelerator
Laboratory, Box 500, Batavia, IL 60510-0500, USA}

\maketitle

\begin{abstract} We study the interaction between monopoles and
embedded domain walls in a $O(3)$ linear sigma model. We discover that
there is an attractive force between the monopole and the wall. We
provide evidence that after the monopole and domain wall collide, the
monopole unwinds on the wall, and that the winding number spreads out
on the surface. These results support the suggestion by Dvali et
al. (hep-ph/9710301) that the monopole problem can be solved by
monopole-domain wall interactions.

\end{abstract}

\section{Introduction} 

Monopoles are predicted in all grand unified field theories which are
spontaneously broken to yield the Standard Model of strong, weak and
electromagnetic interactions \cite{tHooft,Polyakov}. More generally,
monopoles arise whenever a symmetry group $G$ is broken to a subgroup
$H$ such that the vacuum manifold ${\cal M}$ (which in the case of a
simply connected group $G$ is ${\cal M} = G/H$) has nontrivial second
homotopy group, i.e. $\Pi_2({\cal M}) \neq 1$). Zel'dovich and
collaborators \cite{Zel} and Preskill \cite{Preskill} realized that
there is a serious conflict between grand unified field theories (GUTs)
and standard cosmology: GUT symmetry breaking in the early Universe
would produce an over-abundance of monopoles which would overclose the
Universe by many orders of magnitude.

The most popular solution of the monopole problem is to invoke a
period of inflation \cite{Guth} after GUT symmetry breaking. Inflation
will exponentially dilute the number density of monopoles. Since
inflation is likely generated by a sector of the theory that is only
weakly coupled to standard model fields, it is quite possible that the
GUT symmetry is not broken after inflation, and hence that the
monopole problem re-emerges. Within the context of explosive energy
transfer during inflationary reheating \cite{TB} (``preheating"
\cite{KLS}) it is also possible that monopoles may be generated during
reheating
\cite{KLS2,Tkachev,KK,PS}.

There are other possible solutions of the monopole problem, e.g. the
Langacker-Pi mechanism \cite{LP} in which the symmetry is partially
restored at some intermediate time (which leads to the decay of the
heavy monopoles), or the anti-unification scenario \cite{TT,DMS} in
which the GUT symmetry is never restored. These mechanisms, however,
require a substantial amount of particle physics fine tuning.

Recently, Dvali et al. \cite{DLT} have proposed a new solution to the
monopole problem. They consider a theory in which, in addition to
monopoles, domain walls are also present. It was conjecttured that the
domain walls would sweep up monopoles and antimonopoles and, if a
local attractive force between monopole and domain wall exists, the
monopole charge would diffuse onto the domain wall. This would then
lead to an annihilation between monopole and antimonopole charge, thus
solving the monopole problem.

To provide evidence in support of the Dvali et al. mechanism it is
crucial to study the following issues:

\begin{enumerate}
\item{} Is there an attractive force between a monopole and a domain wall? 
\item{} If 1 is correct, then is there scattering, transmission or
does the monopole unwind upon contact with the domain wall?
\end{enumerate}

We study these issues using a scalar field theory with an internal
symmetry of $O(3)$ breaking spontaneously to $ SO(2) $.  This theory
admits both embedded domain walls and global monopoles.  The monopoles
are topologically stable, whereas the embedded walls are not stable
\cite{embedded}. However, we find that the embedded walls in our
simulation have a sufficiently long decay time so that the interaction
between a monopole and the wall can be studied.

Based on our numerical study, we find that there is an attractive
force between the monopole and the wall. Furthermore, we provide
evidence that, upon contact, the monopole unwinds on the wall, with
the winding number spreading out on the surface.

Ideally, one would like to study the interaction of local monopoles
and topologically stable domain walls. For computational reasons we
have decided to begin with an analysis of global monopoles interacting
with embedded walls, a problem involving many fewer fields. We are
interested in local forces rather than global confiugurations, and we
give qualitative arguments that the results of this study will also
apply to the case of local monopoles interacting with stable domain
walls.

Domain walls and Monopoles are ubiquitous in physics. They arise not
only in particle physics, but also in condensed matter physics and
even in string/M theory. The interaction of defects of different types
are important for shedding light on various topics in these branches
of physics (see e.g. \cite{Trodden}). For example, M2 branes and D2
branes are domain wall defects in M-theory and type IIB string theory
respectively.  Recently, it has been suggested that D-branes might
play a significant role in pre-big bang cosmology \cite{Riotto}.
Hence, if there exist important non-trivial interactions between
M-theory domain walls and topological defects like magnetic monopoles,
then such effects may alter the current pre-bang cosmological scenario
in the context of M-theory.

Due to the nonlinearities, interactions between hybrid soliton
networks are not well understood and hard to study analytically (see
e.g. \cite{Vilenkin}). Hence we resort to a numerical study. In the
following section we introduce our model, the equations of motion, and
the defects which the system admits. In Section 3 we outline the
numerical method and code tests. In Section 4 we present and discuss
our results. We use natural units in which $\hbar = c = 1$.

\section{The Model}

We consider a $O(3)$ linear sigma model which is spontaneously broken
to $SO(2)$ by the choice of the ground state. This is achieved by
making use of a ``Mexican hat" potential. The action is
\be 
S \, = \, \int d^4x[ {1 \over 2} \partial_\mu \phi^a \partial^\mu \phi^a
- {1 \over 4} \lambda(\phi^a \phi^a -\eta^2)^2] \, ,   \label{eq1} 
\ee
where summation over $a$ is implied ($a = 1,2,3$). The equations of
motion resulting from (\ref{eq1}) are
\be 
\partial_\mu\partial^\mu \phi^a \, = \, \lambda\phi^a(\phi^b\phi^b -
\eta^2) \label{eom}  
\ee  
 
The vacuum manifold of this theory is
\be
{\cal M} \, = \, {{O(3)} \over {SO(2)}} \, \simeq \, S^2
\label{eq3}
\ee
and has nontrivial second homotopy group. Hence, topologically stable
global monopoles exist. A spherically symmetric monopole centered at
the origin of the coordinate system is described by the following
field configuration:
\be
\phi \, = \, (\phi_1, \phi_2, \phi_3) = f(r,t) (\frac{x}{r},
\frac{y}{r}, \frac{z}{r})\eta \, . 
\label{mon} 
\ee  
where $r^2 = x^2 + y^2 + z^2$ and where the profile function $f(r,t)$
obeys $f(r,t) = f(r) \rightarrow 0$ as $ r \rightarrow 0$ and $f(r,t)
= f(r) \rightarrow 1 $ as $ r \rightarrow \infty$.  A reasonable
ansatz for the profile function at the initial time $t_0$ is \be
f(r,t_{o}) \, = \, (1-e^{- \frac{-r}{r_c}}) \, , \label{eq5}
\ee
where $r_c$ is the core radius of the monopole which is determined by
balancing gradient and potential energies, which yields $r_c \sim
\lambda^{-1/2} \eta^{-1}$.

Since the vacuum manifold ${\cal M}$ is simply connected, there are no
topologically stable domain walls. However, we can easily construct
embedded walls, static but unstable solutions of the equations of
motion. An embedded wall is constructed by picking an axis through the
origin in field space (which intersects ${\cal M}$ in two points), and
considering the corresponding domain wall solution. Choosing the
y-axis we obtain:
\be
\phi_2 = \eta \tanh[r_c^{-1}(y-y_0)], \,\,\, \phi_1 = \phi_3 = 0 \, ,
\label{wall} 
\ee
where $y_0$ is the y coordinate of the center of the wall (the wall
lies parallel to the x/z plane). In our simulation, the wall decays
sufficiently slowly to allow the study of the monopole-wall
interaction.

For the field configuration corresponding to a monopole at the origin
of the coordinate system and an embedded wall located at $y = y_0$, we
choose the following product ansatz:
\be 
\phi_{mdw} = f(r) (\frac{x}{r}, A \tanh[r_c^{-1}(y-y_0)]\frac{y}{r},
\frac{z}{r})\eta \, , \label{mdw} 
\ee 
where $A = \pm 1$. In the limit of large separation, i.e. $r_c^{-1}
y_0 \gg 1$, the field configuration near the origin of the coordiante
system is almost identical to the monopole configuration (\ref{mon}),
and near $y = y_0$ (and for small values of $x$ and $z$), the
configuration coincides with the embedded wall (\ref{wall}).

The topological charge associated with $\Pi_2({\cal M})$ (which in our
case is ${\cal Z}$) is the winding number. This number quantifies how
many times the field configuration $\phi(x)$ wraps around the vacuum
manifold as $x$ ranges over a sphere $S^2$ in coordinate space. Hence
the winding is defined as the homotopy class of the map $\hat{\phi}=
\frac{\phi^a}{|\phi|}$ from coordinate space to the vacuum manifold,
known as isospace: $\hat{\phi} :S_{space}^{2} \rightarrow S_{iso}^2$.

For our O(3) theory, the winding number is:
\be
N= \frac{1}{8\pi} \oint
dS_{k}\epsilon^{ijk}\epsilon_{abc}\hat{\phi^a}\partial_i\hat{\phi^b}
\partial_j\hat{\phi^c}. 
\label{wind} 
\ee  
This integral computes the flux of topological charge through a closed
two-surface.  It follows immediately that the winding number of the
monopole configuration (\ref{mon}) is one, and the winding of the
embedded domain wall (\ref{wall}) considered in our investigation is
zero if the surface for which the winding is evaluated is taken to be
a box surrounding part of the wall.  We can use this information to
test the accuracy of our code which calculates the winding.

Since we consider the time evolution of the winding over the entire
coordinate space, useful information is obtained from the winding
number density.  Using Stokes' theorem in Eq. \ref{wind} by performing
a total derivative on the surface flux in the integrand, the winding
number becomes:
\be N= \frac{1}{8\pi} \int
d^{3}x\epsilon_{abc}\epsilon^{ijk}\partial_i\hat{\phi^a}\partial_j
\hat{\phi^b}\partial_k\hat{\phi^c}.
\label{wind2}
\ee
By visualizing the evolution of the integrand in eq \ref{wind2}
information about the topological charge density over the whole space
is provided.  Therefore, one is able to track the topological charge
of the monopole and understand its non-trivial dynamics.  It is more
challenging to track the charge by studying the surface flux alone,
since one has to know ahead of time where the winding is and choose a
correct surface of integration.  Since, it is difficult to predict the
trajectory and the locality of the monopole charge as it interacts
with the domain wall it is useful to see the entire charge
distribution over the target space; the winding density.  On the other
hand, the monopole winding density is initially a delta function since
$\epsilon_{abc}\epsilon^{ijk}\partial_i\hat{\phi^a}
\partial_j\hat{\phi^b}\partial_k\hat{\phi^c} =\delta(r)$.  On the
lattice the delta function is ill defined, and this poses a problem
with the conservation of the winding number as it is represented on
the lattice.  So it is good to use the winding density information to
determine the location of the winding of the monopole, and then to use
the surface integral to study the winding locally once we know where
it is roughly located.  Since both the surface and volume integrals
have their strengths and weaknesses, we utilize both methods.

\section{Numerical Analysis}

We analyzed the evolution of the field configuration described by Eqs.
(\ref{eom}) on a $100\times 100
\times100$ cubic lattice employing the staggered leapfrog method with
second-order spatial and temporal differencing. Furthermore, the
Courant stability condition was imposed: $\frac{c\delta t}{\delta x}
\leq \frac{1}{\sqrt 2}$. The fields were rescaled such that $\eta =
1$. The coupling constant $\lambda$ was set to 1. Since the gradient
energy was relatively large across the core of the domain wall, the
spatial and temporal resolution was increased by choosing 
$\delta t = 10^{-2}$ and $\delta x = 10^{-1}$.  

To check the accuracy and stability of the code, the total energy was
tracked and it was constant to better than $1\%$ over the entire
running time of the code for the monopole configuration. For the
combined monopole and domain wall, the energy conservation was not as
precise (possibly due to edge effects), but we checked our results by
repeating the calculation on a $200^3$ grid and saw no significant
differences.  In order to test the accuracy of the winding number
algorithm, the winding was calculated for cubic surfaces of differing
sizes around the monopole core.  Fifth order differencing was employed
for the winding calculation.  The topological charge was equal to 1
within 0.1\%; It was independent of numerical noise, edge effects, and
the size of the cubic surface.  In addition, the winding density was
calculated but it was not properly conserved.  This is expected
because the winding density for a monopole is a delta function, which
is not reproducible on the lattice.  However, on qualitative and
physical grounds one can still trust the results for the winding
density since the delta function is smeared out over four grid cubes,
independent of the number of total grid cubes in the whole space,
which demonstrates that the computer's representation of the delta
function is independent of the resolution imposed.

The boundary conditions must be chosen with care. Dirichlet boundary
conditions are inconsistent with the presence of a net winding number
(or embedded winding) in the box. With periodic boundary conditions,
coordinate space becomes the torus $T^3$, and therefore the winding
over the surface of the box must vanish. Another way to see the
problem with periodic boundary conditions is to realize that the
asymptotics of the hedgehog configuration (\ref{mon}) is inconsistent
with periodic boundary conditions. A similar problem arises for
Dirichlet boundary conditions. However, by smoothing out the field
configuration at the edge of the box, the hedgehog configuration
(\ref{mon}) can easily be made consistent with Neumann boundary
conditions. Hence, we considered Neumann boundary conditions, i.e. we
set the derivatives of the fields at all edges of the cube to zero to
smooth out the fields at large distances. Note that Neumann boundary
conditions are consistent with the topology of coordinate space being
${\cal R}^3$, and with the winding number evaluated for the surface of
the box being an integer.

With periodic boundary conditions, it is possible to construct a field
configuration which looks like a monopole close to the center of the
box, but the fields must be distorted compared to the configuration
(\ref{mon}) at the edge of the box, and this introduces a compensating
negative winding number localized near the edge of the the box. The
positive and negative winding numbers can annihilate, and thus the
local monopole configuration is unstable, as we verified in our
simulations.

With Neumann boundary conditions, we first studied the stability of
our code by studying a single monopole configuration. Figure 1 shows
the value of $|\phi| = {\sqrt{(\phi^a \phi^a)}}$ (vertical axis) in
the x-y plane at four different time steps. The initial width $r_c$
was chosen to be smaller than the critical value $\lambda^{-1/2}
\eta^{-1}$ (see Section 2). As a consequence, we observe a ringing of
the monopole. However, even by starting the fields away from their
stable configuration, no instability is observed. Hence, as it should
be, the hedgehog configuration (\ref{mon}) is observed to be a stable
configuration in our simulations.

\begin{figure}[htbp] 
\begin{center}
\begin{tabular}{cc}
\epsfysize=5cm 
\epsfbox{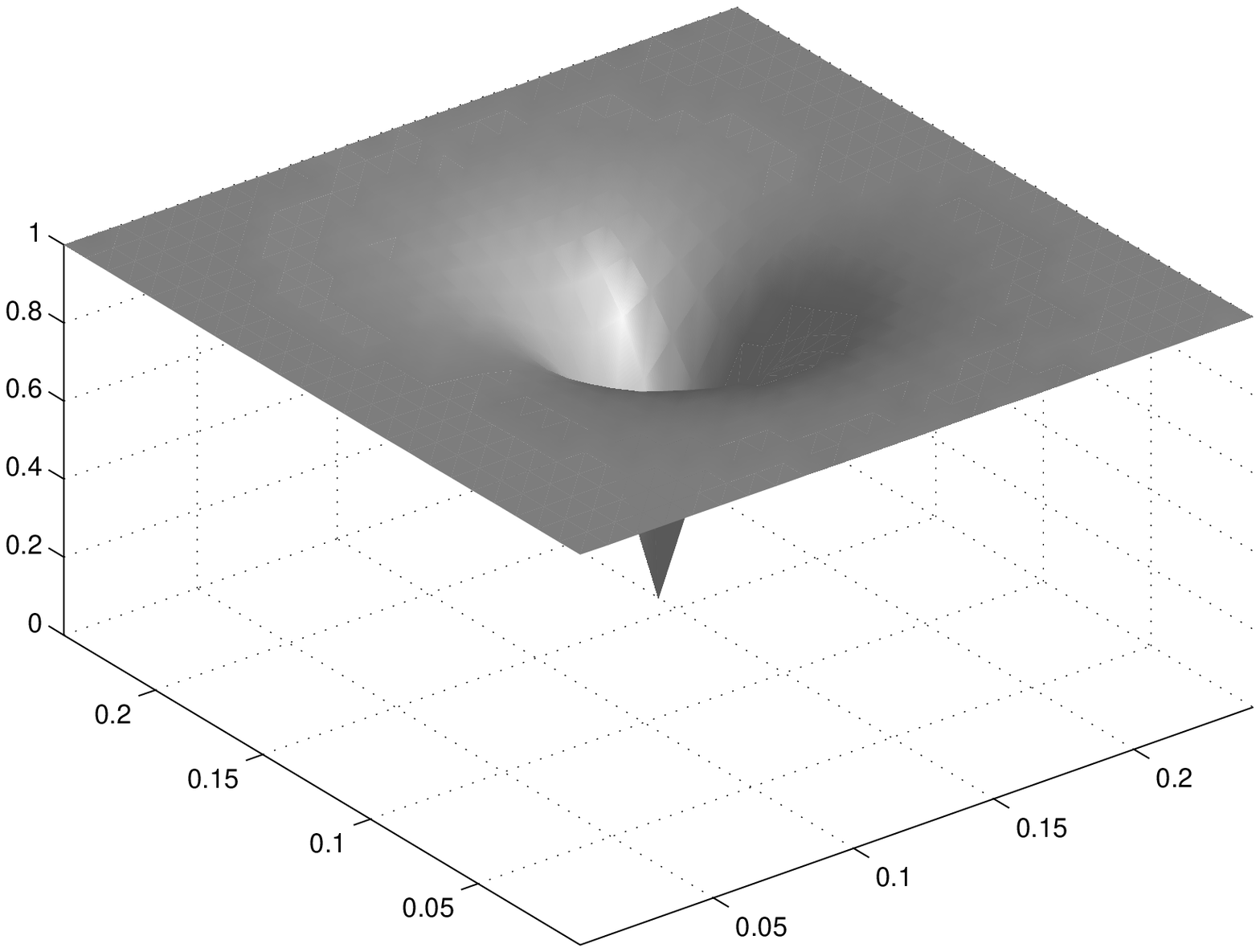} &~~~~
\epsfysize=5cm 
\epsfbox{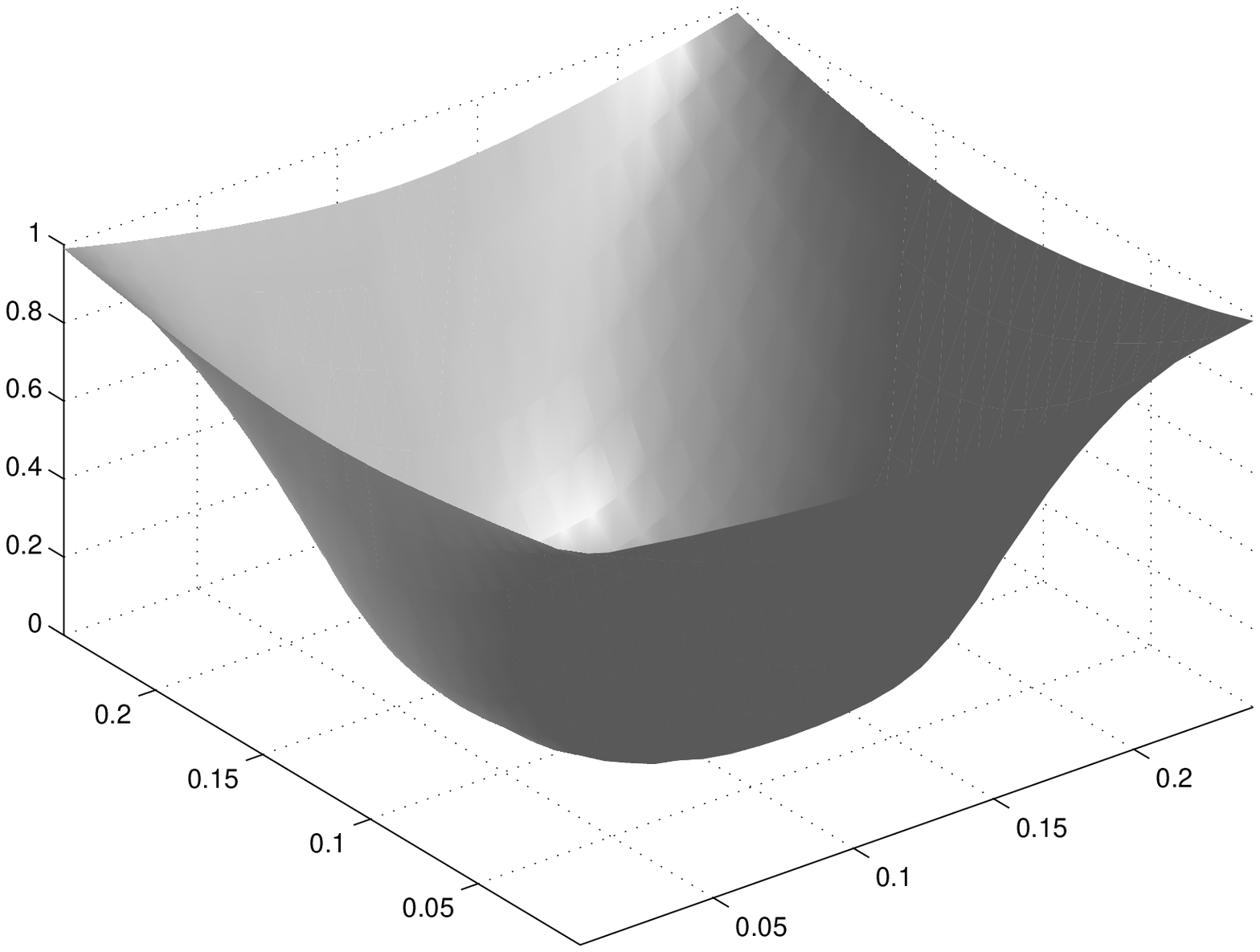} \\
\epsfysize=5cm 
\epsfbox{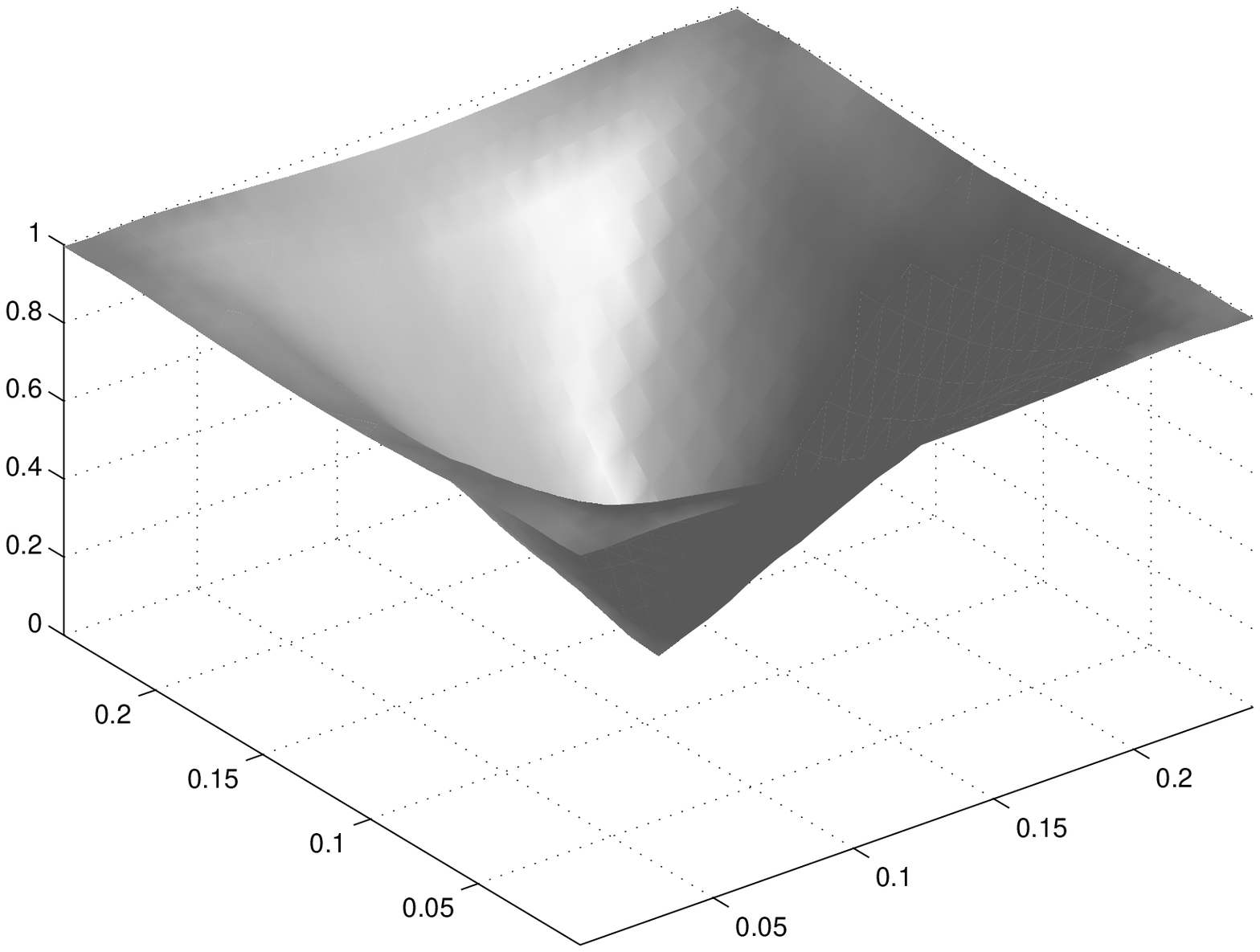} &~~~~
\epsfysize=5cm 
\epsfbox{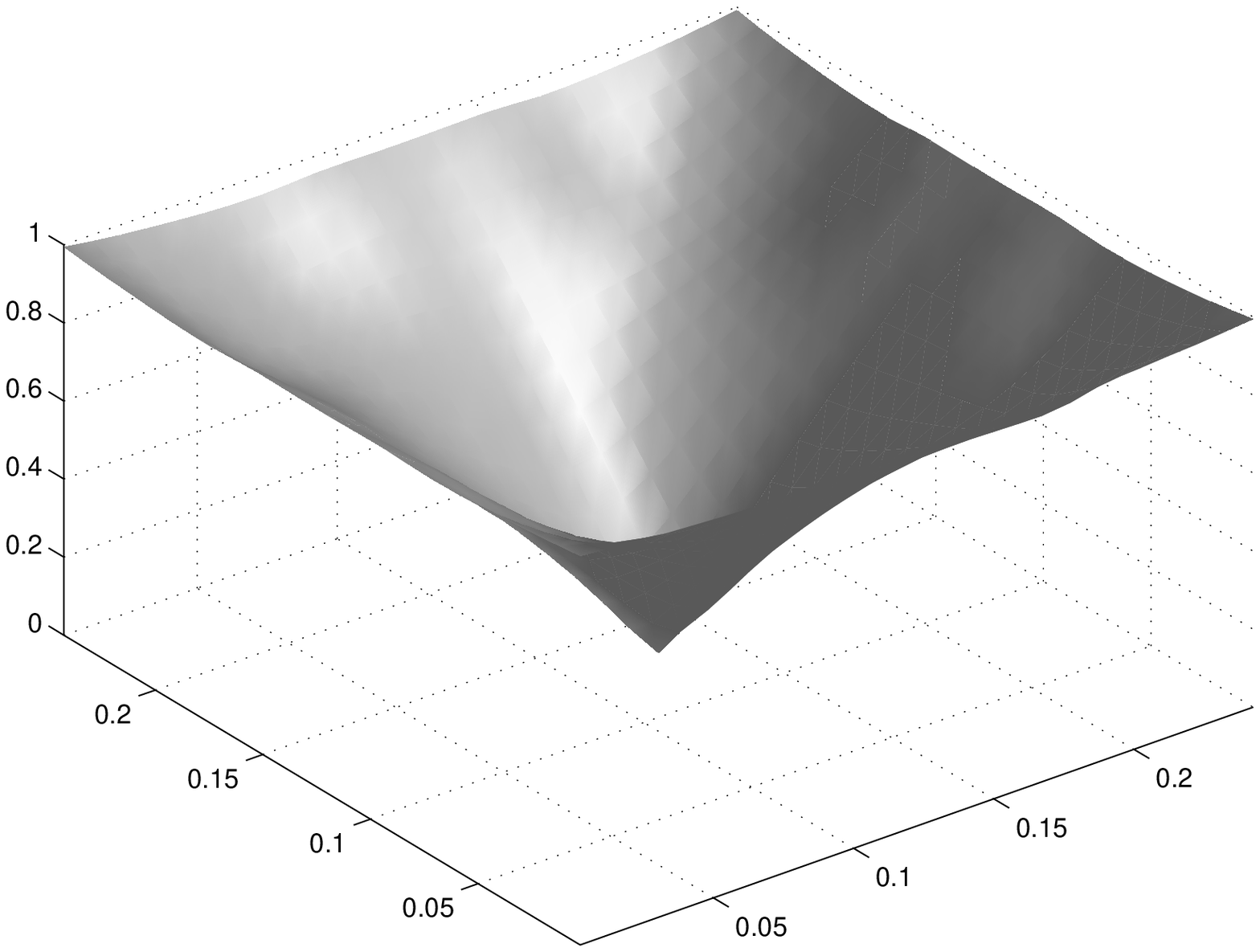} \\
\end{tabular}
\caption[]{Evolution of a monopole configuration with winding number
$N = 1$. The vertical axis represents $|\phi|$, the horizontal plane
is the x-y plane. Although the winding number is constant, the
monopole core oscillates in width.}  
\end{center}
\end{figure}

\section{Results and Discussion}

\subsection{Stability of an Embedded Domain Wall}

Five years ago, Barriola, Vachaspati and Bucher \cite{Bucher} showed
that a new class of non-topological defects, {\it embedded defects},
can exist in many field theories, in particular in the electroweak
theory. They are solutions of the field equations. In a theory with
vacuum manifold ${\cal M}$, embedded defects correspond to topological
defects of a theory whose vacuum manifold is a submanifold of ${\cal
M}$. Typically, they correspond to configurations in which some of the
fields of the full theory are set to zero by hand. As was shown
analytically in \cite{Bucher}, embedded defects are unstable to
perturbations of the field configuration.  Our embedded wall
(\ref{wall}) is an example of an embedded defect.  The domain wall
(\ref{wall}) is an embedded defect and is unstable precisely because
the $Z_2$ which is broken is a subgroup of O(3).
        
Our embedded domain wall (\ref{wall}) is a two dimensional membrane
that interpolates between two disjoint vacua $\phi_{\pm}$ which are
out of phase by $180^o$. Figure 2 shows the time evolution of the
energy density of the field configuration. By tracking the energy
density we found that, independent of the spatial and temporal
resolution and of the thickness of the wall core, the embedded wall
(\ref{wall}) splits up into two walls which attain equal and opposite
velocities. This splitting can be understood as follows. Initially, at
the center of the wall the field is at a local maximum of the
potential energy density. The forces which render the embedded wall
unstable want to bring the field into a vacuum state. The forces are
strongest in the core of the wall, and hence it is along the core that
the fields will move away from their original values fastest. As this
happens, the potential energy is transformed into kinetic and gradient
energy. The induced field gradients will then act on the points to the
right and to the left of the core and will start to pull them into the
potential valley with them. Thus, gradient/kinetic energy waves are
generated which propagate away from the original embedded wall
position with the speed of light. Note that in the case of planar
symmetry these new ``walls" carry zero monopole winding number
(because the field configuration will lie on a one-dimensional subset
of the vacuum manifold). Note that this behavior of the unstable
embedded wall occurs also for periodic boundary conditions.

The splitting of the embedded wall configuration will make the
description of monopole-wall interaction more complicated. In the
presence of a monopole, the planar symmetry of the wall gets
destroyed, and then the two gradient/kinetic energy walls may well
acquire a fractional winding number, as will be illustrated below. The
tracking of the winding number density of the monopole-wall
configuration becomes more difficult.  Note that the splitting of the
embedded wall occurs before the fields feel the reflective boundary
conditions. Note also that the coherence of the domain walls survives
long enough to be able to study the monopole-wall interaction.

\begin{figure}[htbp]
\begin{center}
\begin{tabular}{cc}
\epsfysize=5cm
\epsfbox{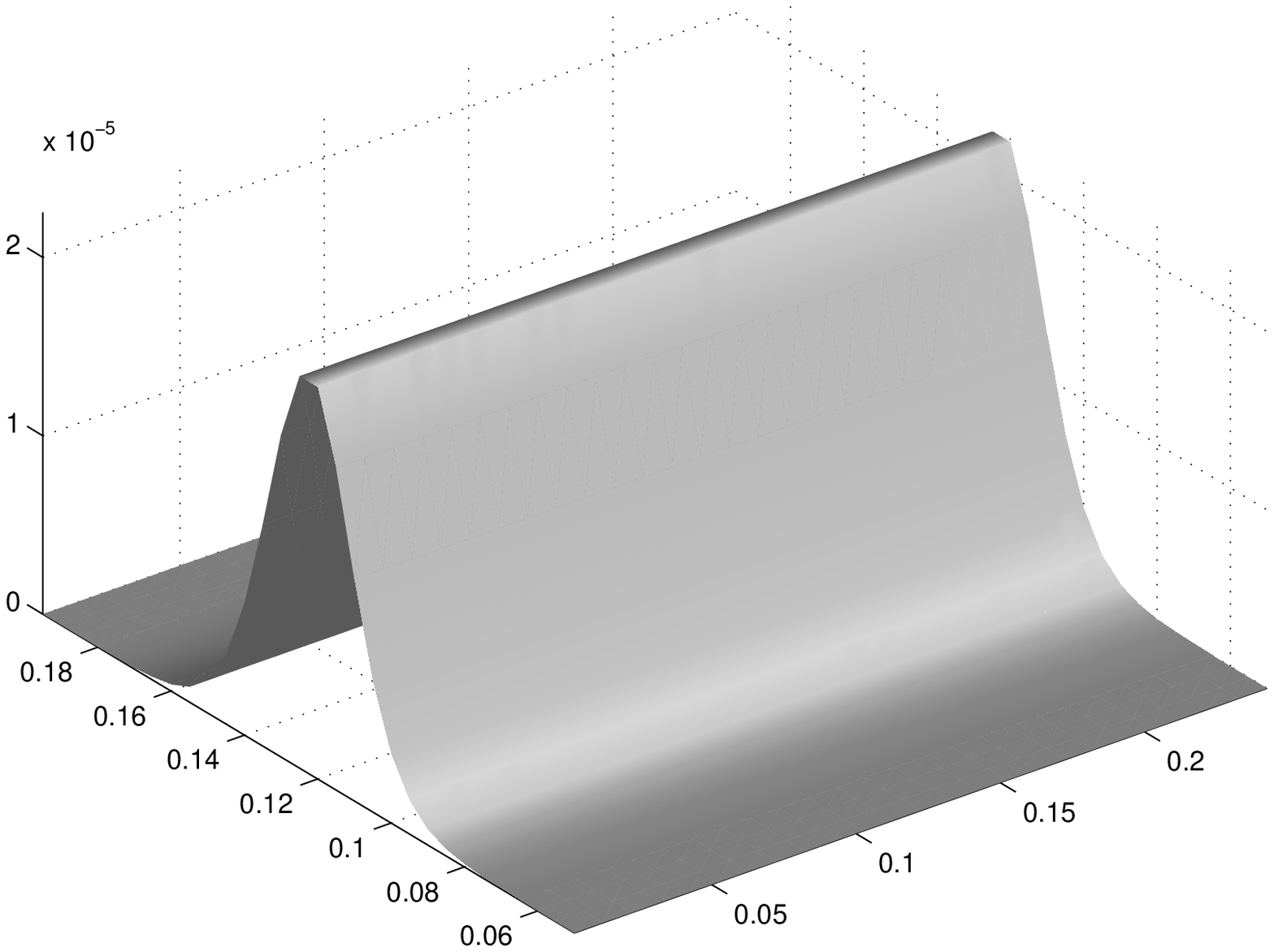} &~~~~
\epsfysize=5cm
\epsfbox{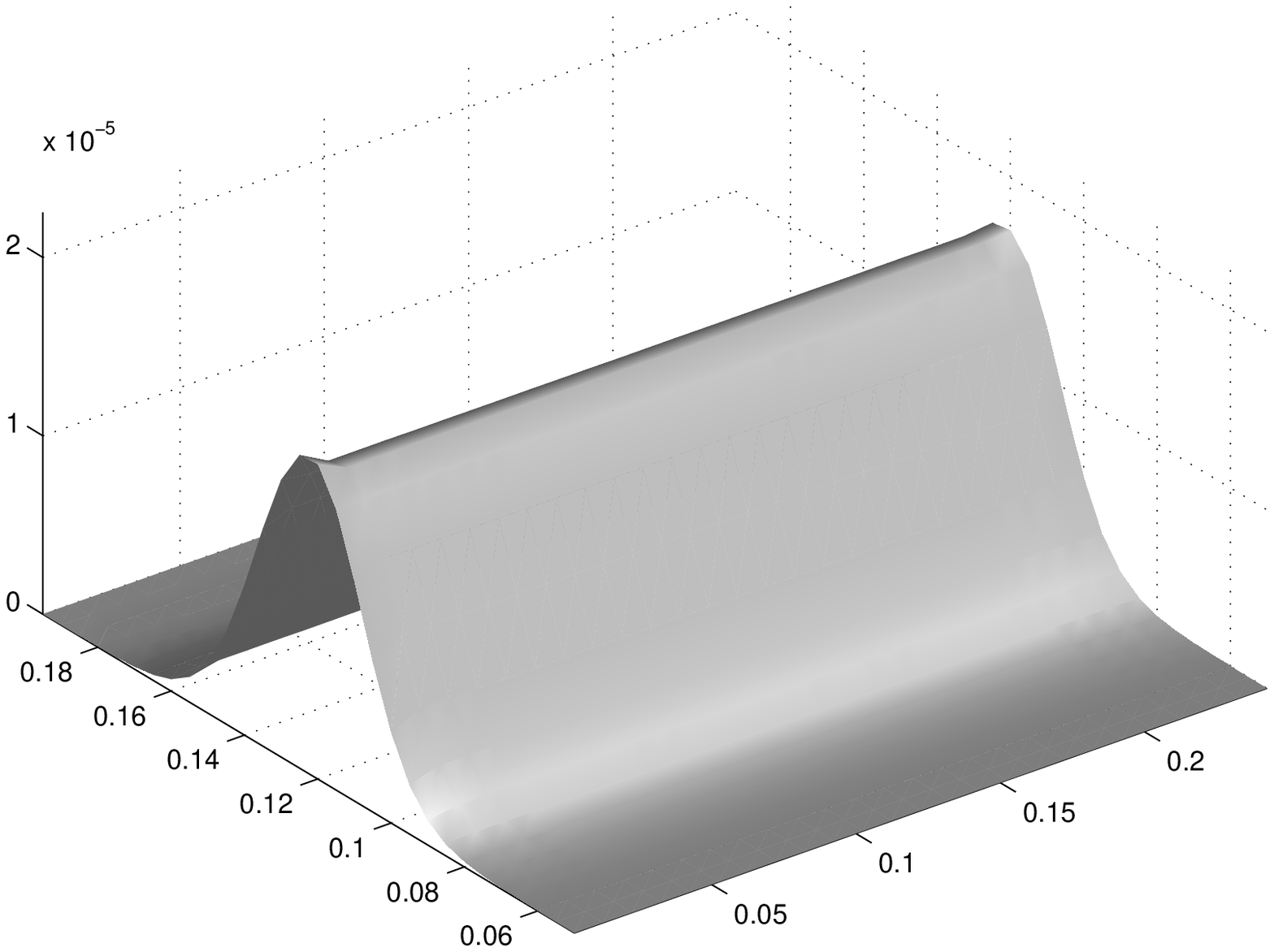} \\
\epsfysize=5cm
\epsfbox{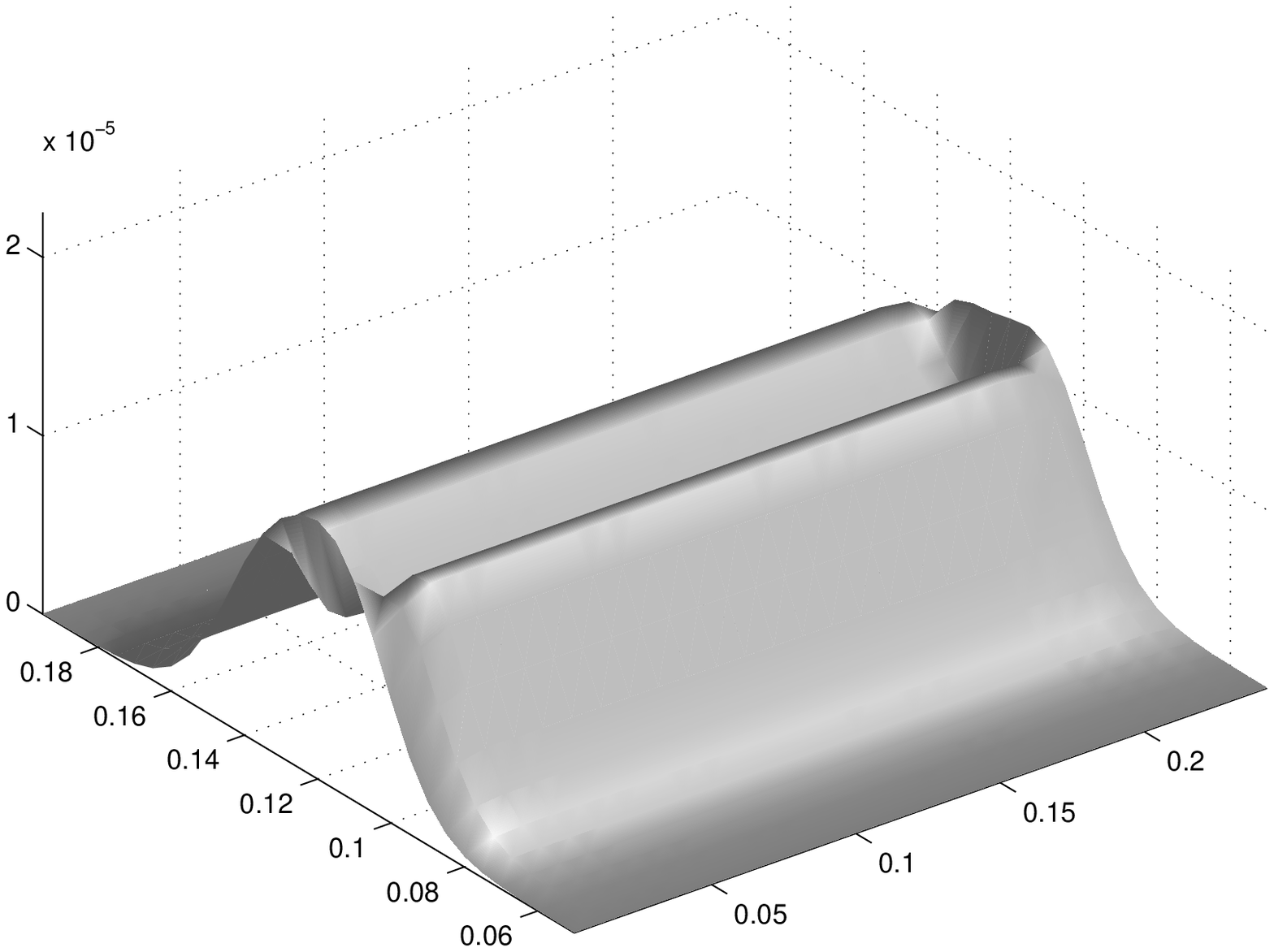} &~~~~ 
\epsfysize=5cm
\epsfbox{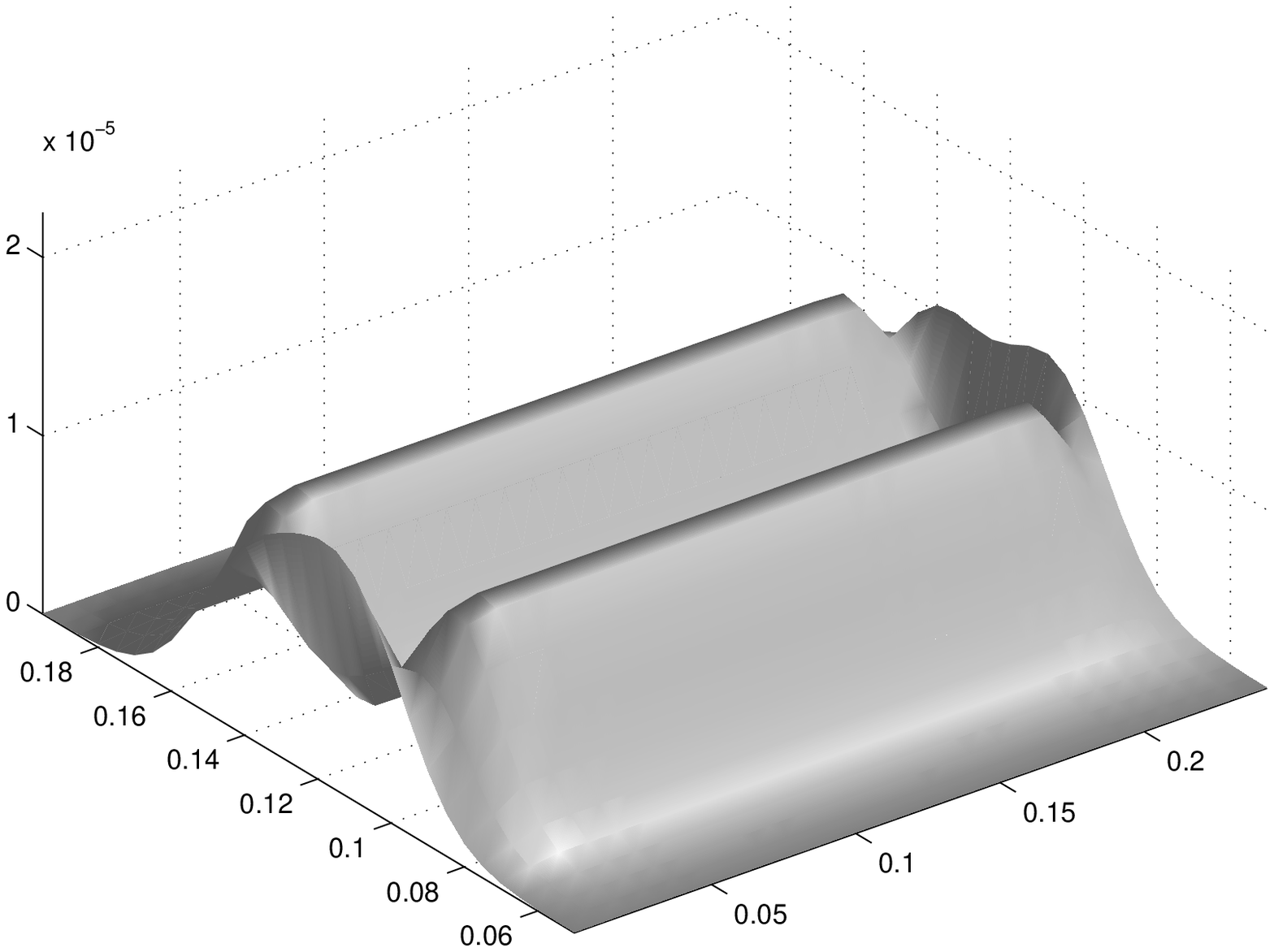} \\
\end{tabular}

\caption[]{The evolution of a free domain wall.  Eventually the
initial wall splits into two energy walls with equal magnitude but
oppositely oriented velocities.}  
\end{center}
\end{figure}
 
\subsection{Attractive Force between the Monopole and Domain Wall}

We studied the monopole-wall interaction by starting with the static
configuration of (\ref{mdw}) corresponding to the product ansatz of a
monopole at rest and an embedded wall at rest (with $A = -1$). By
observing the evolution of the energy density as a function of time in
the x-y plane (Figure 3) it is possible to partially address the
nature of the domain wall- monopole interaction.

{F}rom the numerical results shown in Figure 3 it follows that there
is an attractive force between the monopole and the wall. If the
distance between the monopole and the wall is large, then the core of
the monopole evolved in an identical way to the free monopole
evolution which was previously discussed in the previous
section. However, if the distance is small, then this is not the
case. Beginning in the second time step shown in Figure 3, it becomes
clear that the presence of the wall will distort the monopole
configuration (which, in the absence of the wall, was stable). The
energy density peak of the monopole splits, with one peak being pulled
towards the domain wall. This peak which at later times carries most
of the monopole energy merges with the wall. As can be seen in the
last graph, the energy is absorbed by the wall. There is no
reflection. The wall persists and continues to sweep across the grid
until it reflects off the edge of the target space. We have verified
that the attractive force exists for both signs of $A$ in
(\ref{mdw}). Furthermore, there is an identical attractive force
between the embedded wall and the antimonopole.

One way to understand the origin of the attractive force, is to
perform an adiabatic analysis and consider the total gradient energy
as a function of the separation $s$ between the wall and the
monopole. The total gradient energy decreases as $s$
decreases. Therefore, since the field configuration tends to minimize
its gradient energy, there will be an attractive force. The decrease
in the total gradient energy can be seen as follows. Consider the
configuration (\ref{mdw}) with $A = -1$. By sketching the field
configuration (see Figure 4) it is easy to convince oneself that as
long as $s > r_c$, it is mostly the gradient energy of the wall which
is effected by the separation $s$. Note that while $|\phi|$ is fixed,
only the sign of the field component $\phi_2$ changes as we cross the
embedded wall, and it is only this change which contributes to the
gradient energy. As $s$ decreases, the area of the wall where $\phi_2$
is large also decreases, and hence the total gradient energy
decreases.  The decrease in gradient energy density is not uniform in
$\rho = {\sqrt{(x^2 + z^2)}}$, but it is sharpest near the edges of
the box, as is evident from Figure 3, where it can be seen that the
wall energy rapidly falls off as $x$ approaches the edges of the box.

Although this examination of the energy density of the field
configurations is important in establishing the attractive force
between wall and monopole, it is not enough to verify more subtle
effects, like the possible unwinding of the monopole charge or
monopole creation and annihilation. In particular, it is not possible
to decide if the second peak in the monopole density (the one further
away from the wall) is radiation or if it still carries some winding
number. Topological analyses are neccessary to probe these issues.

\begin{figure}[htbp]
\begin{center}
\begin{tabular}{cc}
\epsfysize=5cm
\epsfbox{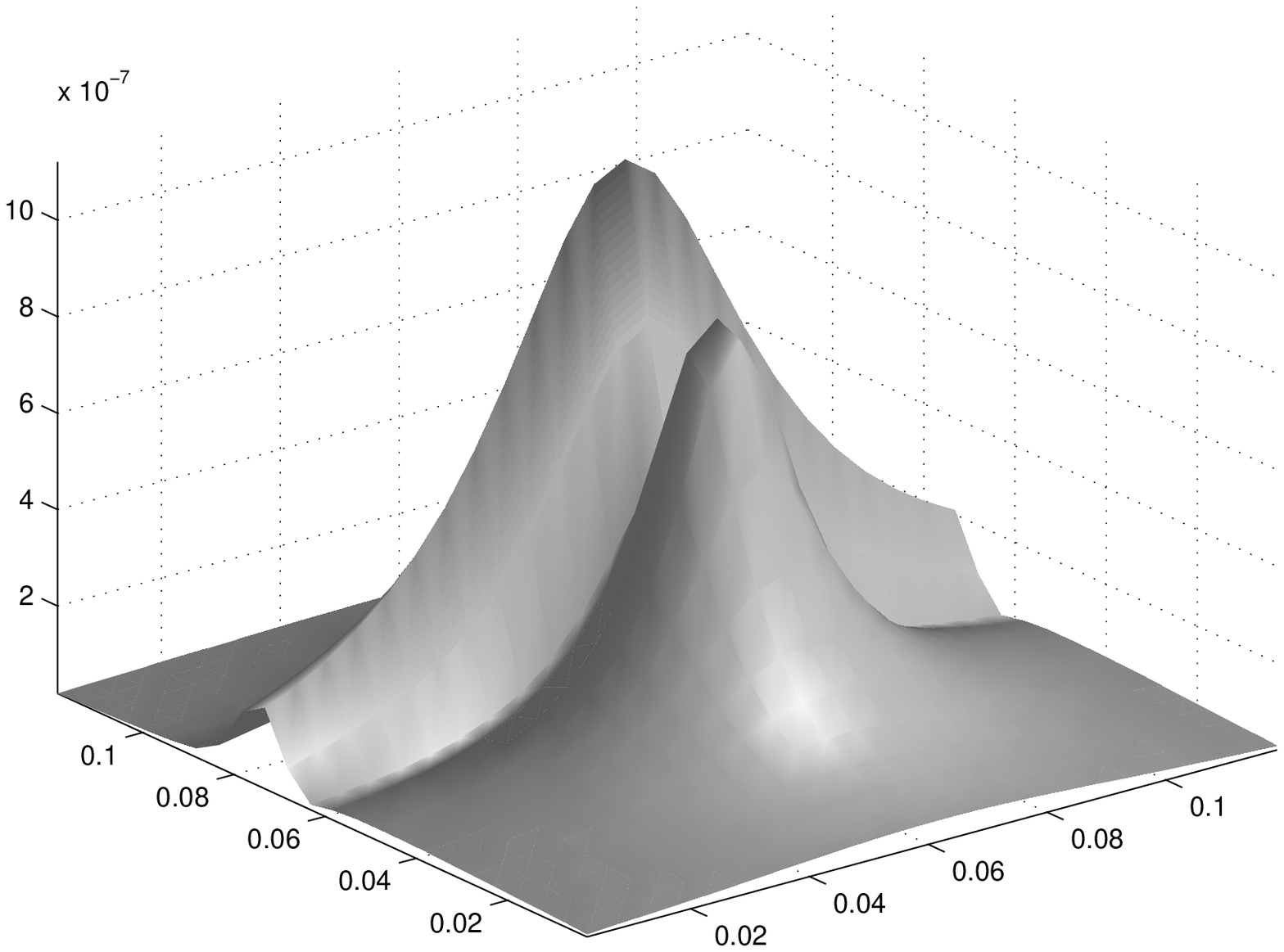} &~~~~
\epsfysize=5cm
\epsfbox{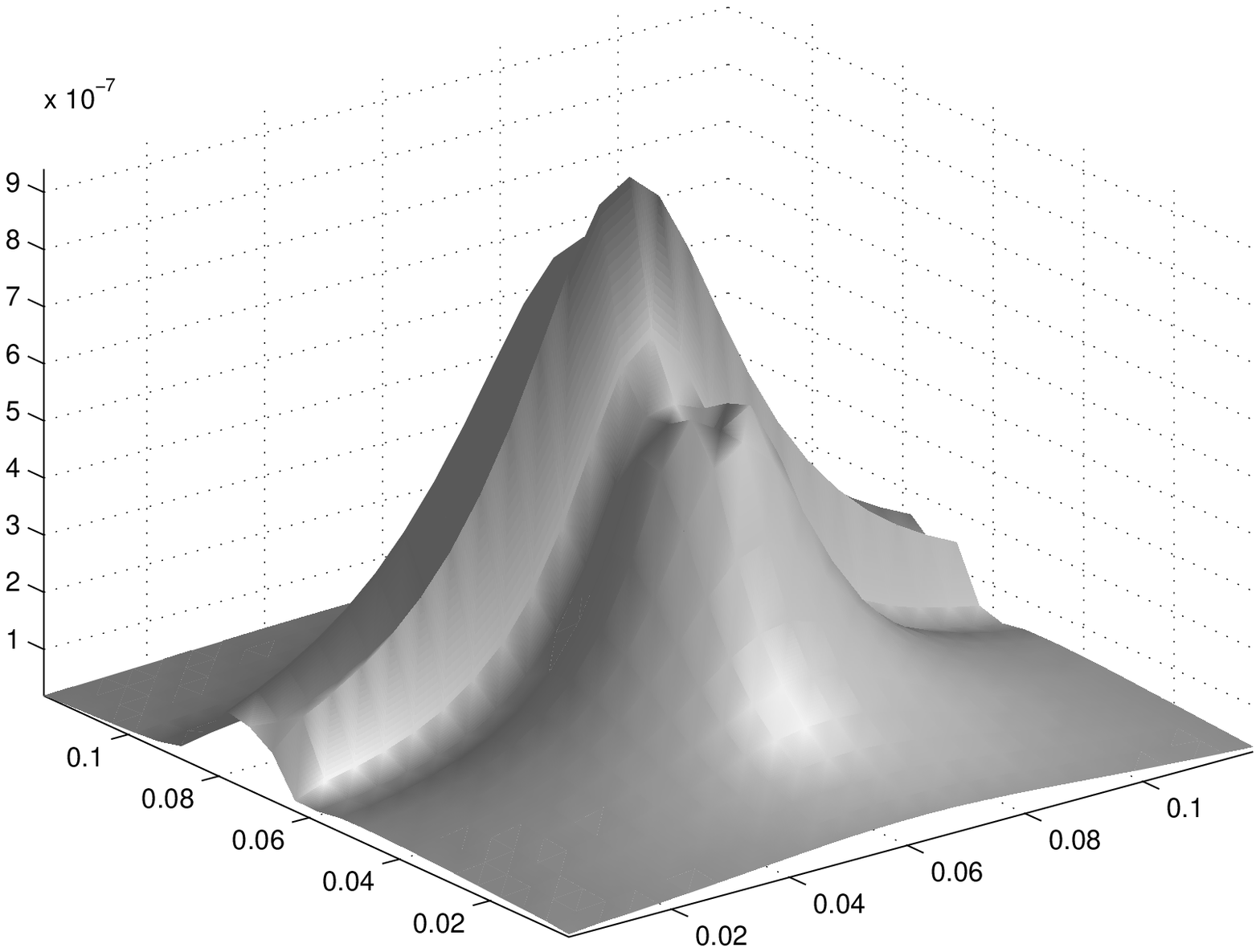} \\
\epsfysize=5cm
\epsfbox{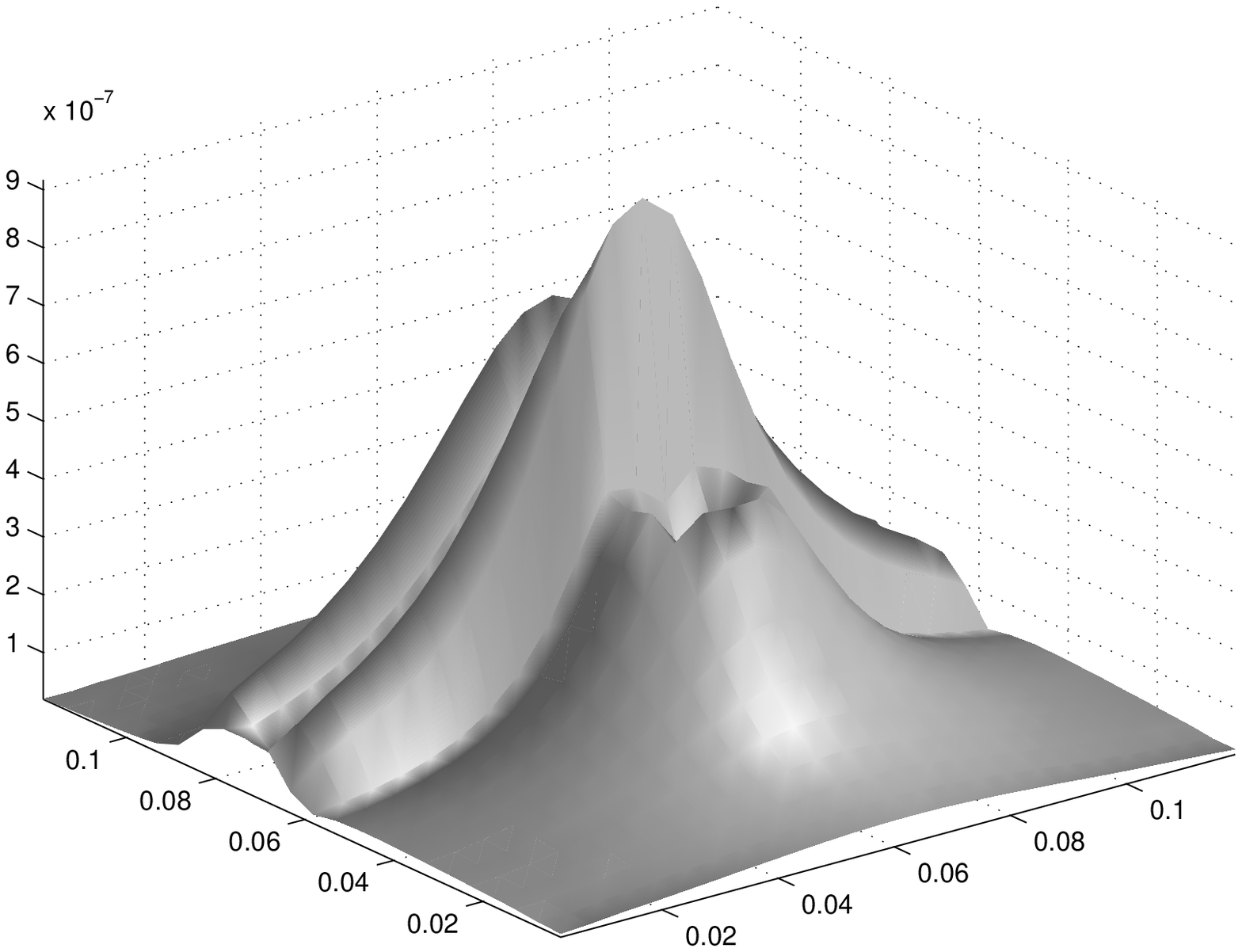} &~~~~
\epsfysize=5cm
\epsfbox{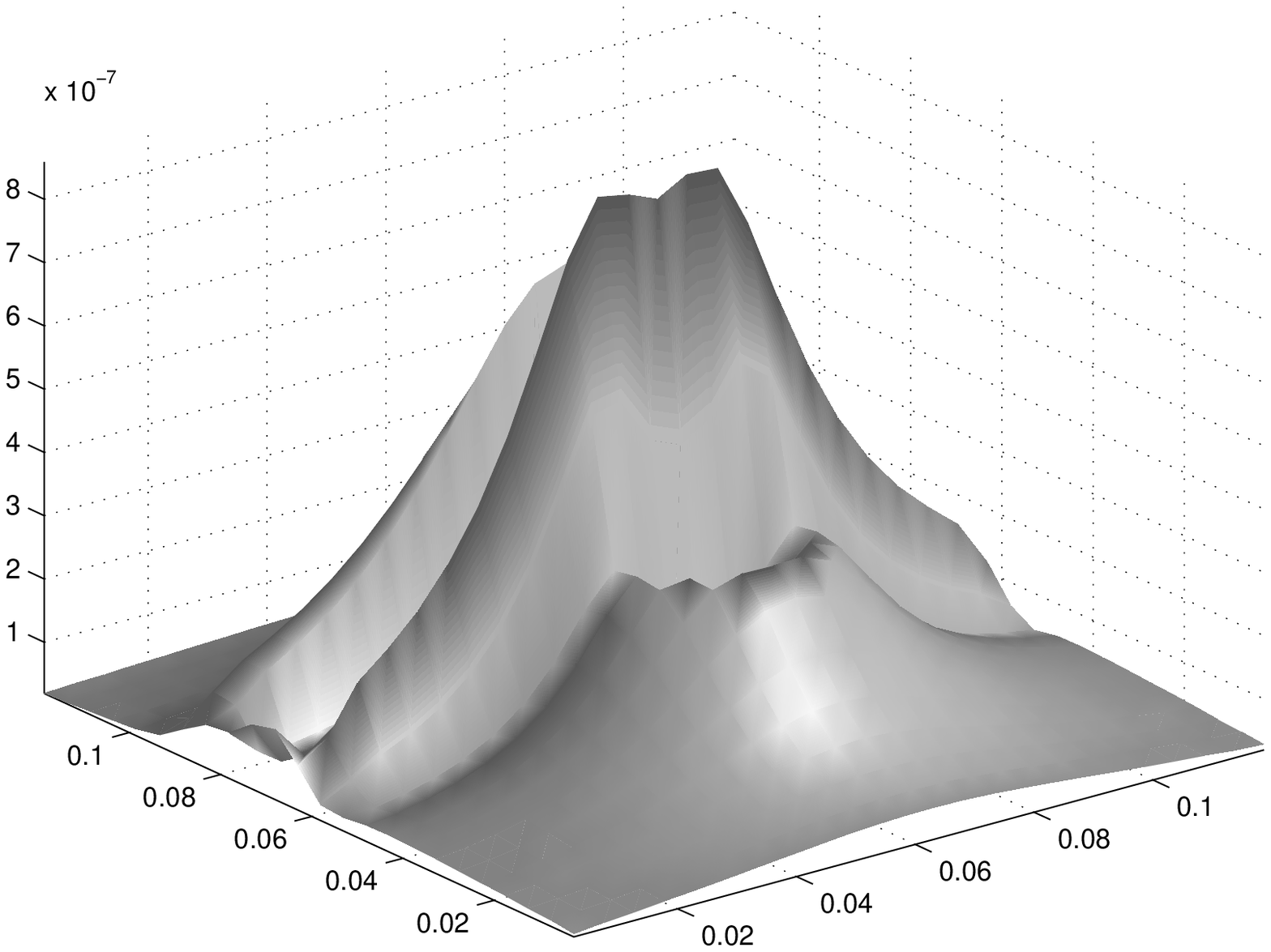} \\
\epsfysize=5cm
\epsfbox{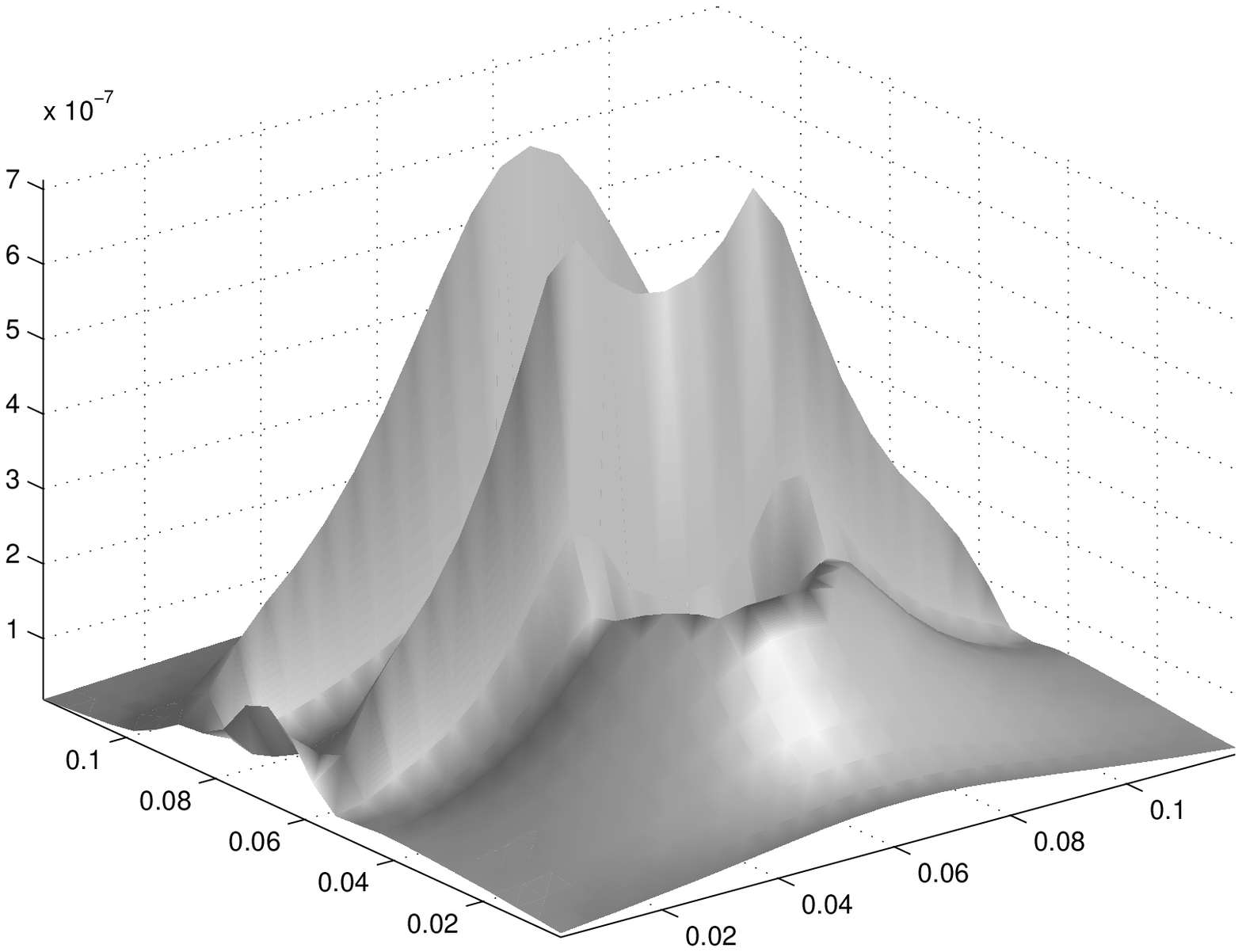} &~~~~
\epsfysize=5cm
\epsfbox{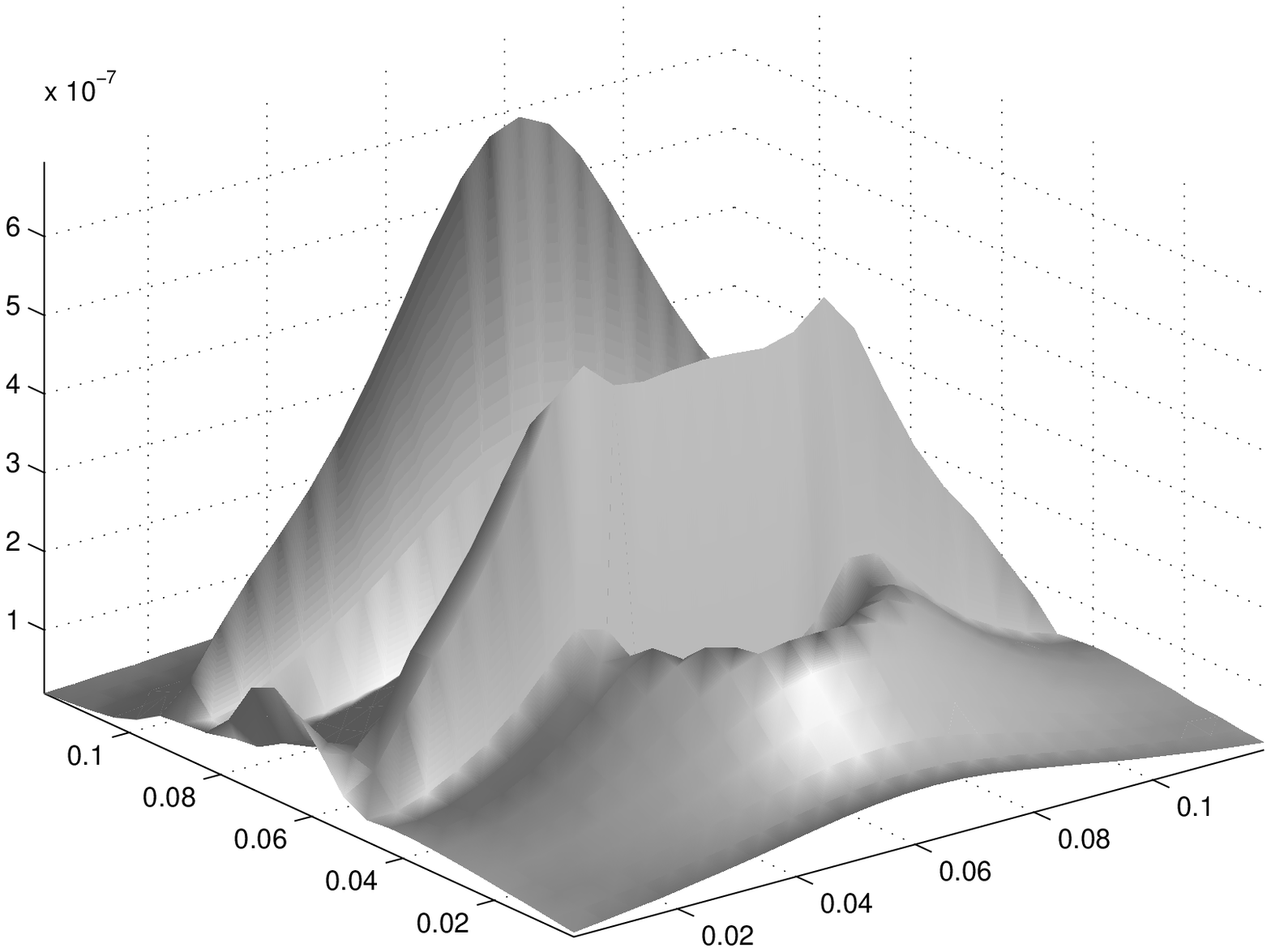} \\
\end{tabular}
\end{center}
\caption[]{Time evolution of the energy density for the initial
configuration (\ref{mdw}) with $A = -1$. This shows the dynamics of
the domain wall - monopole interaction.The monopole experiences an
attractive force towards the domain wall and is stuck to it, and most
of the energy density of the monopole is dispersed onto the wall.}
\end{figure}

\subsection{Unwinding of the Monopole on the Wall}

The question of whether or not the monopole unwinds on the doman wall
surface requires a careful investigation of the winding number
evolution over the entire space.  The protocol for investigating the
winding evolution is as follows:
\begin{enumerate}
\item The winding density over the entire coordinate space is
evaluated in order to determine where the winding number is
localized. 
\item The integrated surface winding is evaluated  for a surface
surrounding the monopole and one surrounding the central section of
the wall, separately. At the same time, as a test of total winding
number conservation, the winding over the surface of the box is
computed.    
\item The winding is tracked on the entire surface of the wall as well
as around a gaussian surface of the monopole to check for unwinding
events. 
\end{enumerate}

\begin{figure}
\begin{center}
\epsfbox{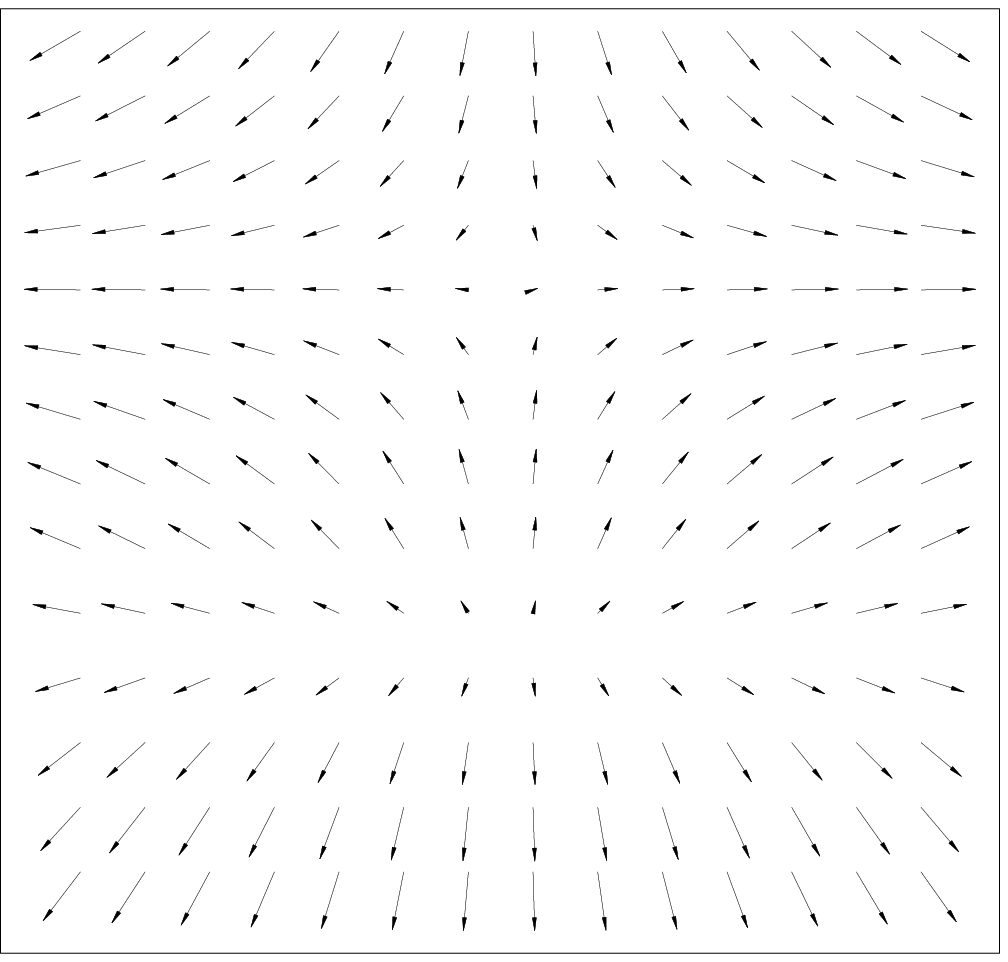}
\epsfysize=10cm
\caption[]{A plot of the phase of $\phi$ in the x-y plane for the
configuration (\ref{mdw}). The domain wall runs horizontally across
the plot, and the monopole is situated beneath it.}
\end{center}
\end{figure}

By tracking the surface winding we confirm that the total winding over
the entire space for the monopole and wall configuration (\ref{mdw})
is conserved. Its value is 0.9. The difference from the naively
expected value of 1 is due to edge effects - where the wall hits the
surface of the box the field is not in the vacuum manifold, and hence
the usual argument for the quantization of the winding number breaks
down. That the total winding is (almost) 1 is to be expected since the
domain wall has zero winding and the monopole has a winding of one.

The evolution of the winding number density over time is shown in
Figure 5. We see that as the monopole approaches the wall, a negative
image winding builds up in the center of the wall. This can be
understood from the sketch of Figure 4. If we track the phase of
$\phi$ ( as a phase in the $\phi_1 - \phi_2$ plane intersecting the
vacuum manifold ${\cal M}$) along a circle in the x-y plane about the
center of the monopole, going in counter-clockwise direction, then the
phase increases monotonically from $0$ to $2 \pi$. This corresponds to
a winding number $N = 1$. However, if we consider a circle in the same
plane of coordinate space surrounding the center of the wall, the
phase decreases monotonically from $\pi$ to $- \pi$, corresponding to
a winding number of $N = -1$.

The emergence of a negative image winding on the wall provides another
way of explaining the attractive force between the monopole and the
wall.  The monopole sees a virtual anti-monopole on the surface of the
wall and is attracted to it.  It is also now obvious why, as is seen
in the third frame of Figure 5, the monopole charge unwinds on the
wall. After the $N = 1$ winding of the monopole and the $N = -1$ image
winding on the wall annihilate, the remnant winding which adds up to
$N = 1$ is left behind on the outer regions of the wall.

Note that an analysis of the winding density in the x-y plane (see
Figure 6) shows that the second peak in the energy density of the
evolved monopole (see Figure 3) carries no winding and thus
corresponds to radiation.

\begin{figure}[htbp]
\begin{center}
\begin{tabular}{cc}
\epsfysize=5cm 
\epsfbox{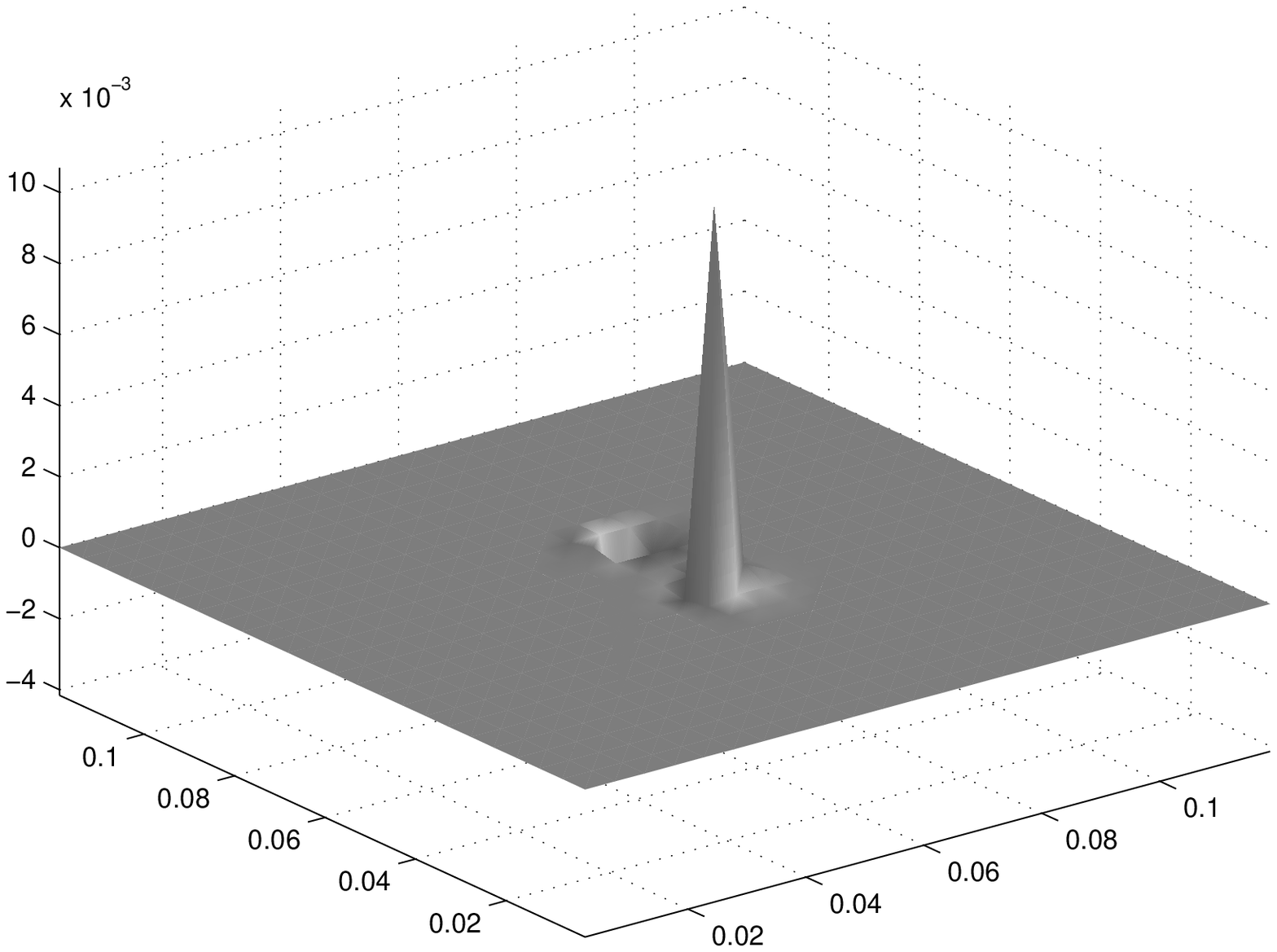} &~~~~
\epsfysize=5cm 
\epsfbox{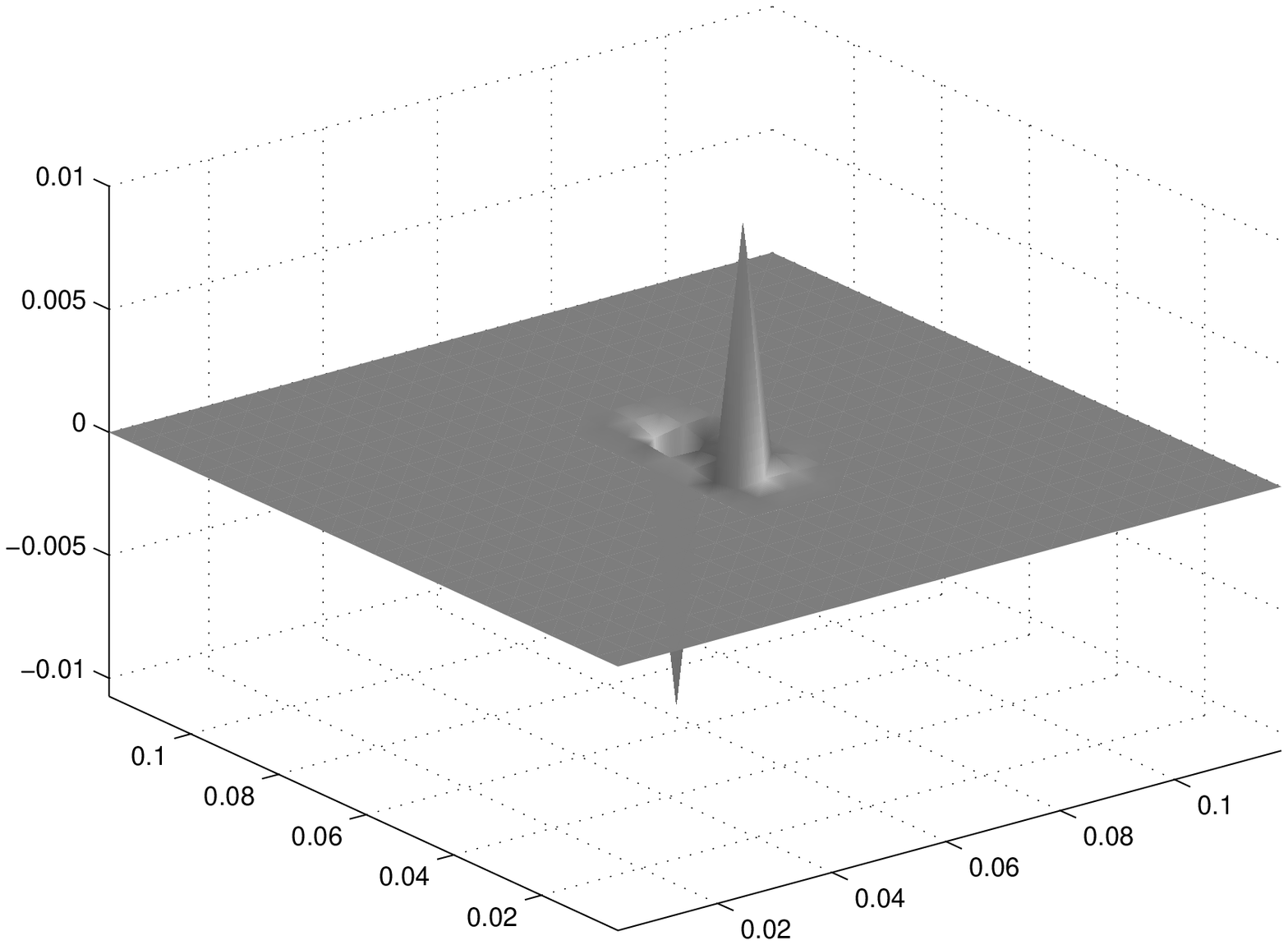} \\
\epsfysize=5cm 
\epsfbox{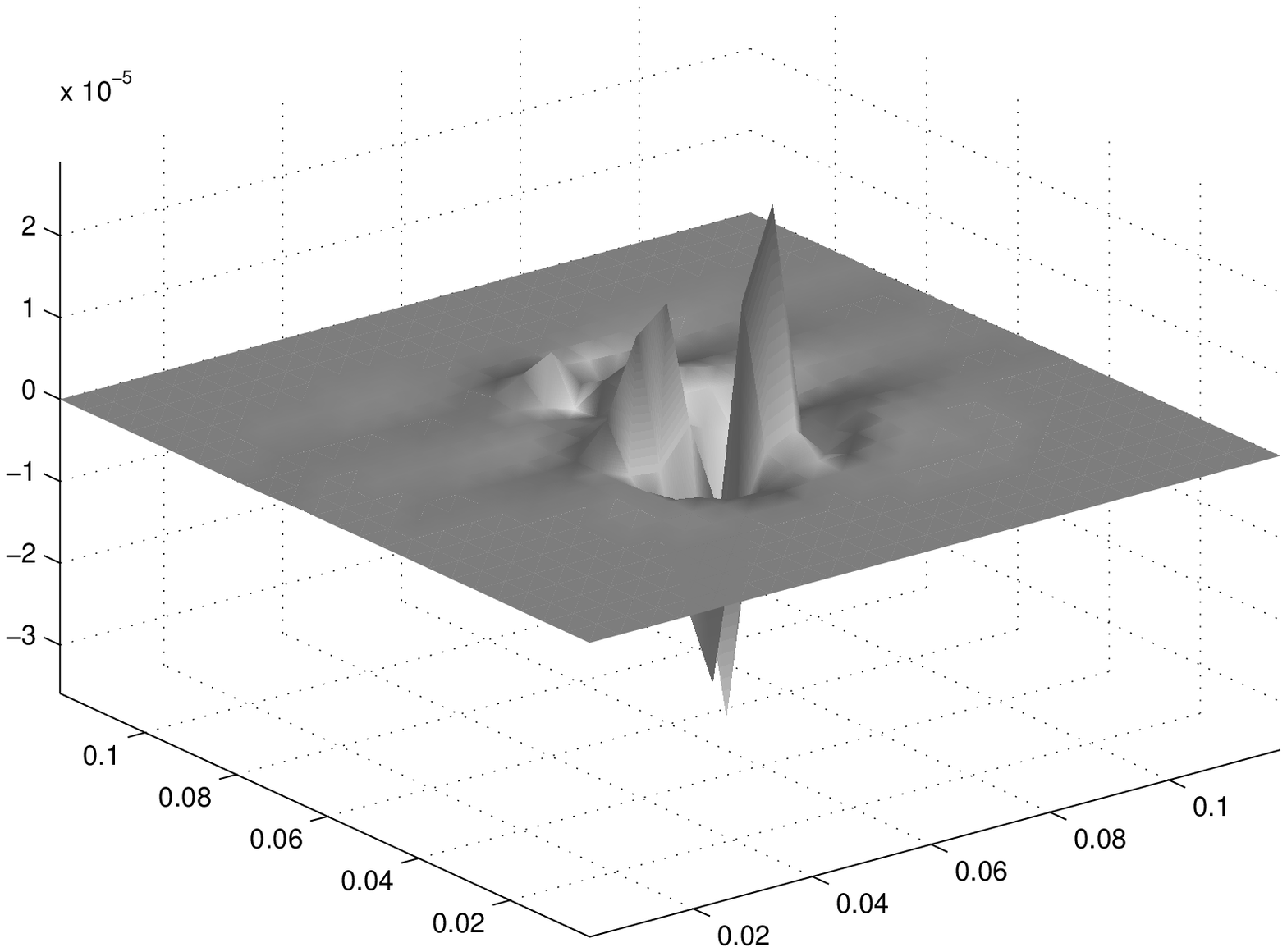} &~~~~
\epsfysize=5cm 
\epsfbox{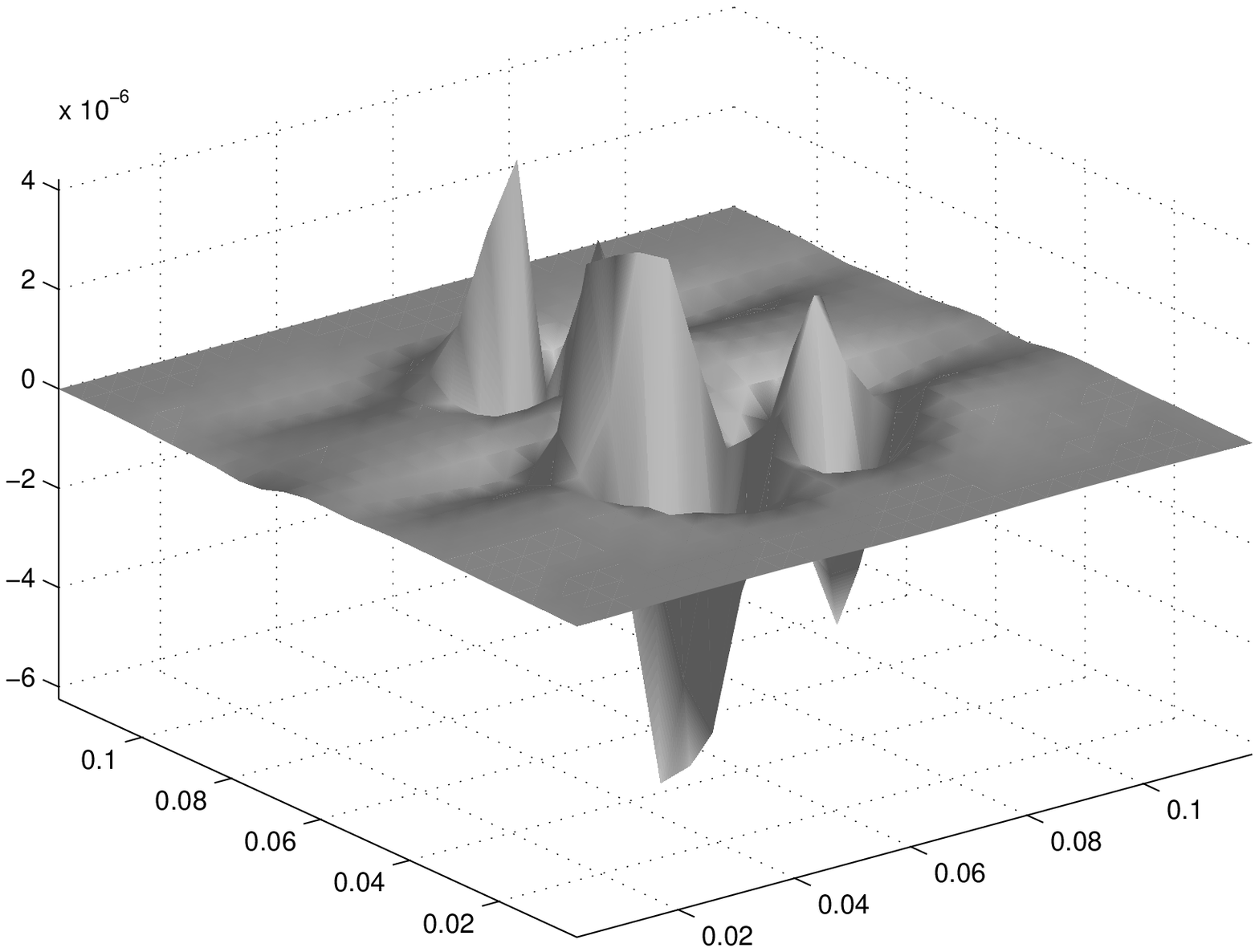} \\
\end{tabular}
\caption[]{Topological charge density as a function of time for a
monopole interacting with a domain wall. The monopole induces an image
charge on the surface of the domain wall.  Eventually, the monopole
unwinds on the surface of the wall. Note that the scale of the $z$
axis differs between the subplots.}
\end{center}
\end{figure}

\begin{figure}[htbp]
\begin{center}
\begin{tabular}{cc}
\epsfysize=5cm
\epsfbox{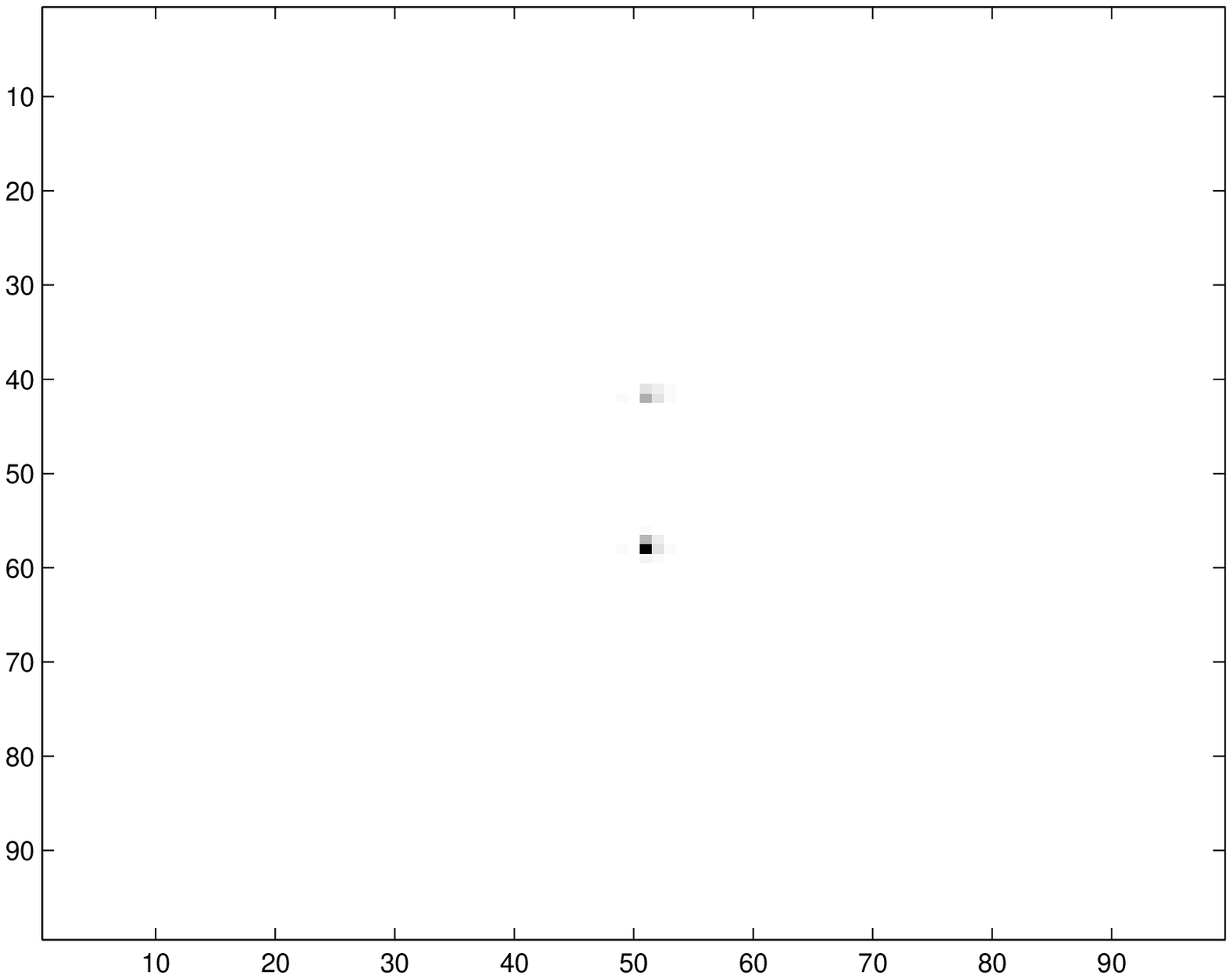} &~~~~
\epsfysize=5cm 
\epsfbox{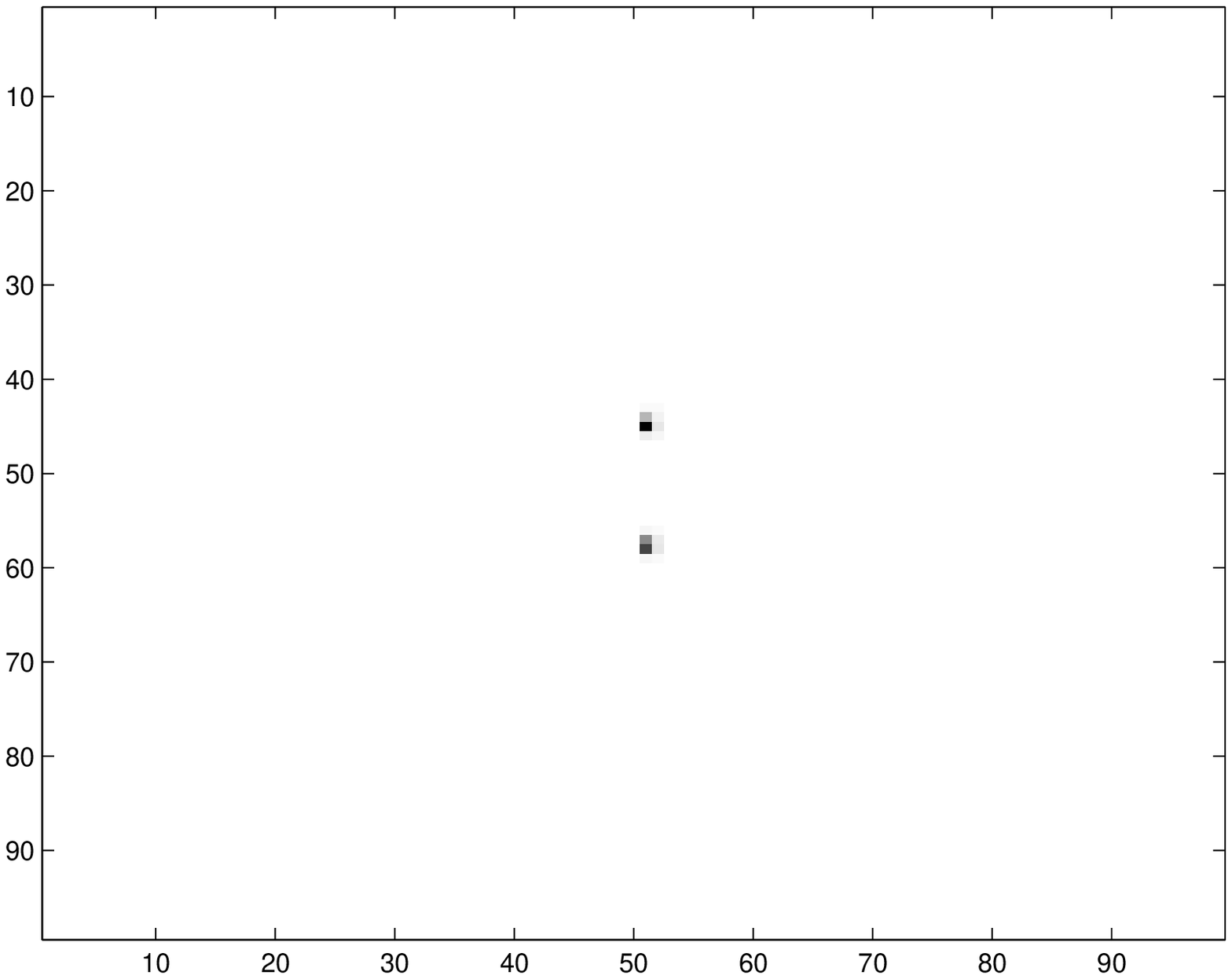} \\
\epsfysize=5cm 
\epsfbox{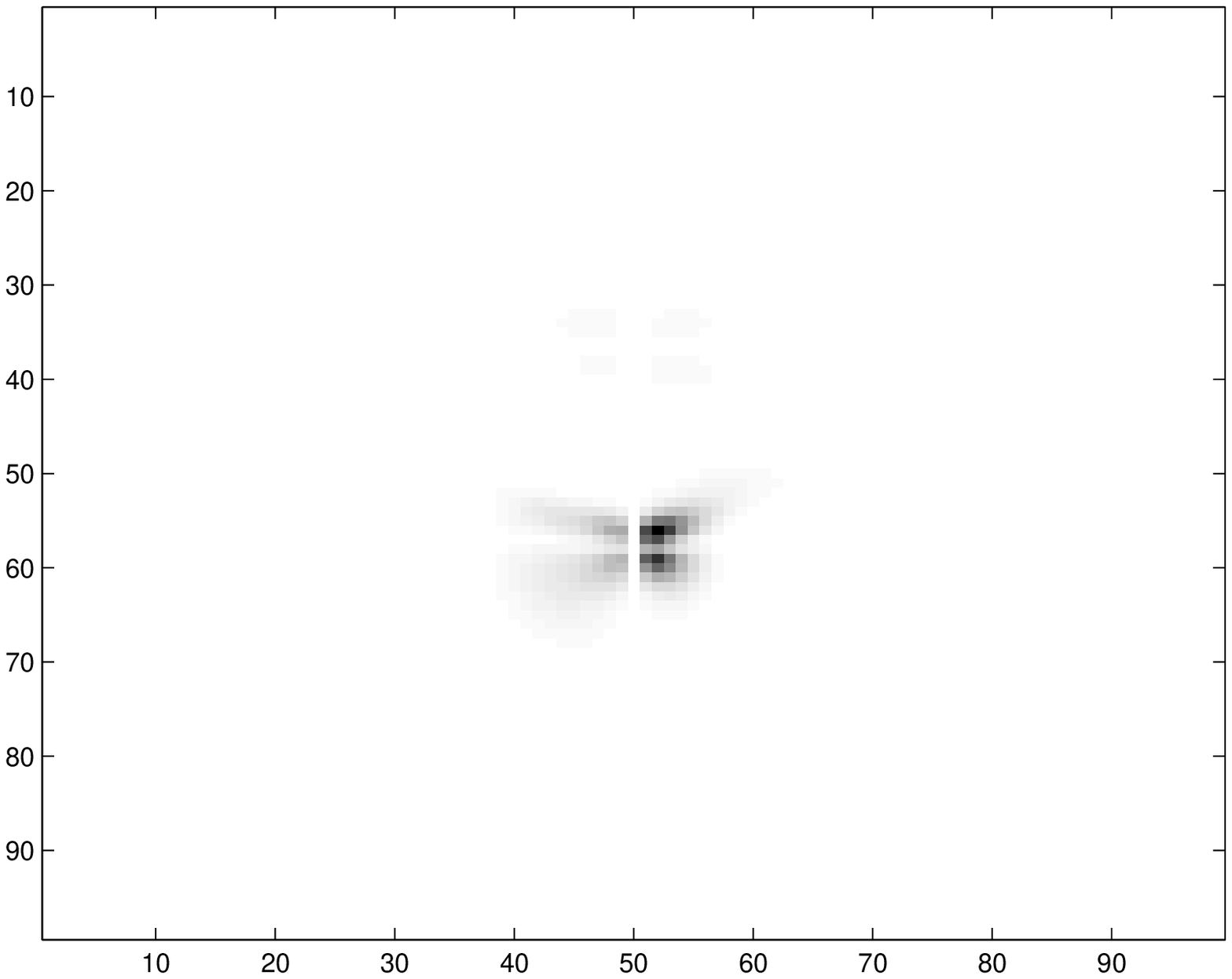} &~~~~
\epsfysize=5cm
\epsfbox{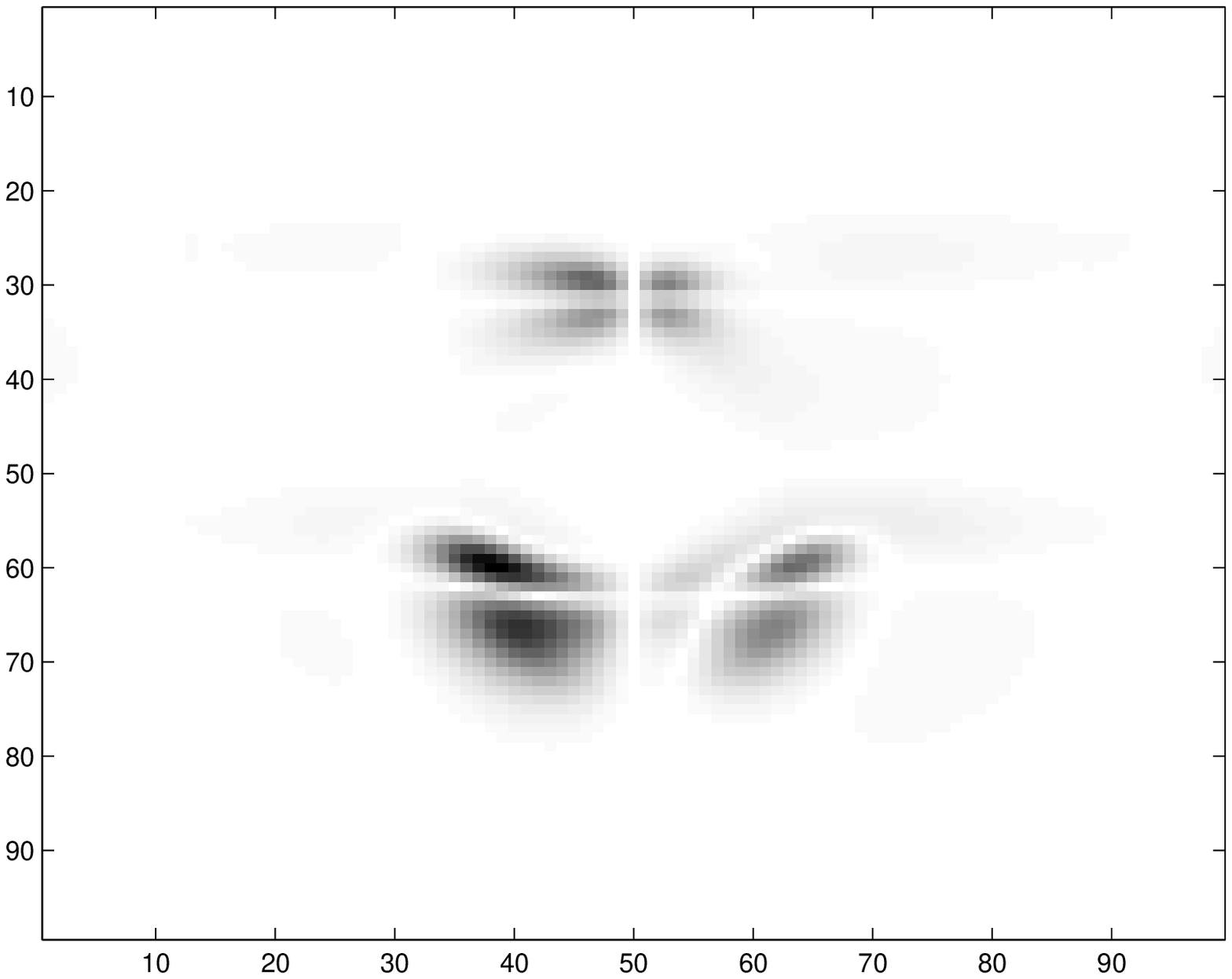} \\
\end{tabular}
\end{center}
\caption[]{Another view of the topological charge density in the x-y
plane for the initial configuration (\ref{mdw}) and its time
evolution. From this set of frames we see that the monopole
winding is attracted to the wall.  The magnitude of the charge density
is plotted. Black represents the maximum absolute value of the winding
density in each subplot.  }
\end{figure}

\section{Conclusions}

We have studied the interaction of a global monopole with an embedded
wall in a $O(3)$ linear sigma model whose symmetry is spontaneously
broken to $SO(2)$ by the Higgs mechanism. Our main conclusions are the
following:
\begin{enumerate}
\item There is an attractive force between the monopole and domain wall. 
\item The monopole charge unwinds on the surface of the domain wall.
\end{enumerate}
We have explained the attractive force as being a result of the
buildup of a negative image winding number on the wall. This image
charge production also explains why the monopole charge unwinds on the
wall upon contact, rather than the monopole simply scattering off the
wall. 

Our results should also carry over to the more complicated but
physically more interesting models with local monopoles and
topologically stable walls as long as monopoles and walls are
(partially) made up of the same fields. Our results thus support the
new solution of the monopole problem proposed by Dvali et
al. \cite{DLT}. 

\section*{ Acknowledgements}
	
It is a pleasure to thank Mark Hindmarsh and Antal Jevicki, who were
helpful at crucial stages.  In particular, we would like to thank
Matthew Parry for his continual useful discussions and suggestions.
Computational work in support of this research was performed at the
Theoretical Physics Computing Facility at Brown University.  This work
has been supported in part by the US Department of Energy under
Contract DE-FG02-91ER40688, TASK A, and also by the DOE and NASA grant
NAG 5-7092 at Fermilab.


\begin{thebibliography}{99} 

\bibitem{tHooft} G. 't Hooft, {\it Nucl. Phys.} {\bf B79}, 276 (1974).
\bibitem{Polyakov} A. Belavin, A. Polyakov, A. Shvarts and
Yu. Tyupkin, {\it Phys. Lett.} {\bf 59B}, 85 (1975).  
\bibitem{Zel} Ya. B. Zel'dovich and M. Yu. Khlopov, {\it Phys. Lett.}
{\bf 79B}, 239 (1978). 
\bibitem{Preskill} J. Preskill, {\it Phys. Rev Lett.} {\bf 43}, 1365 (1979).
\bibitem{Guth} A. Guth, {\it Phys. Rev.} {\bf D23}, 347 (1981)
\bibitem{TB} J. Traschen and R. Brandenberger, {\it Phys. Rev.} {\bf D42},
2491 (1990). 
\bibitem{KLS} L. Kofman, A. Linde and A. Starobinsky, {\it Phys. Rev. Lett.} 
{\bf 73}, 3195 (1994).
\bibitem{KLS2} L. Kofman, A. Linde and A. Starobinsky, {\it
Phys. Rev. Lett.} {\bf 76}, 1011 (1996). 
\bibitem{Tkachev} I. Tkachev, {\it Phys. Lett.} {\bf B376}, 35 (1996).
\bibitem{KK} S. Kasuya and M. Kawasaki, {\it Phys. Rev.} {\bf D56},
7597 (1997). 
\bibitem{PS} M. Parry and A. Sornborger, hep-ph/9805211.
\bibitem{LP} P. Langacker and S-Y. Pi, {\it Phys. Rev. Lett.} {\bf
45}, 1 (1980). 
\bibitem{TT} J. Iliopoulos, D. Nanopoulos and T. Tomaras, {\it
Phys. Lett.} {\bf 94B}, 141 (1980). 
\bibitem{DMS} G. Dvali, A. Melfo and G. Senjanovic, {\it
Phys. Rev. Lett.} {\bf 75}, 4559 (1995). 
\bibitem{DLT} G. Dvali, H. Liu, and T. Vachaspati, hep-ph/9710301
\bibitem{embedded} M. Barriola, T. Vachaspati and M. Bucher, {\it
Phys. Rev.} {\bf D50}, 2819 (1994). 
\bibitem{Trodden} S. Carroll and M. Trodden, {\it Phys. Rev.} {\bf
D57}, 5189 (1998)   
\bibitem{Riotto} M. Maggiore, A. Riotto, hep-th/9811089.
\bibitem{Vilenkin} A. Vilenkin and E.P.S Shellard, ``Cosmic Strings and    
Other Topological Defects" (Cambridge University Press, Cambridge, 1994).
\bibitem{Bucher} M. Barriola, T. Vachaspati and M. Bucher, {\it
Phys. Rev.} {\bf D50}, 2819 (1994). 

  
\end{thebibliography}
\end{document}